\documentclass[runningheads]{llncs}
\usepackage[T1]{fontenc}
\usepackage{longtable}

\usepackage{graphicx}
\usepackage{hyperref}
\usepackage{color}

\usepackage{subfig}

\begin{document}
\title{Semi-Automatic Labeling and Semantic Segmentation of Gram-Stained Microscopic Images from DIBaS Dataset }
\titlerunning{Semi-automatic Labeling and Semantic Segmentation...}
%
\author{Chethan Reddy G.P.$ ^1 $
\and Pullagurla Abhijith Reddy$ ^1 $
\and Vidyashree R. Kanabur$ ^1 $
\and Dr. Deepu Vijayasenan $ ^1 $ \and Dr. Sumam S. David $ ^1 $ \and Dr. Sreejith Govindan $ ^2 $}
\authorrunning{Chethan Reddy G.P.}
%
\institute{ $ ^1 $ National Institute of Technology Karnataka, Surathkal, Karnataka, India \\
 \email{vidyashreerk1992@gmail.com}\\
\and
 $ ^2 $ Melaka Manipal Medical College(MMMC), Manipal Academy of Higher Education, Manipal, Karnataka, India}
%
\maketitle              
\begin{abstract}

In this paper, a semi-automatic annotation of bacteria genera and species from DIBaS dataset is implemented using clustering and thresholding algorithms. A Deep learning model is trained to achieve the semantic segmentation and classification of the bacteria species. Classification accuracy of 95\% is achieved. Deep learning models find tremendous applications in biomedical image processing. Automatic segmentation of bacteria from gram-stained microscopic images is essential to diagnose respiratory and urinary tract infections, detect cancers, etc. Deep learning will aid the biologists to get reliable results in less time. Additionally, a lot of human intervention can be reduced. This work can be helpful to detect bacteria from urinary smear images, sputum smear
images, etc to diagnose urinary tract infections, tuberculosis, pneumonia, etc. 
\end{abstract}
\section{Introduction}
Bacteria are tiny living microorganisms that are found everywhere. Most of the bacteria are beneficial, and very few are harmful. Bacteria are essential to break down the food into nutrients, boost the immune system, etc. Harmful bacteria may cause diabetes, pneumonia, cancer, respiratory and urinary infections, etc. Detection and classification of bacteria based on their genera and species is essential to diagnose these infections and diseases. \par
Classification of bacteria into various genera and species can be done based on their shape, stain-type, arrangement and size. Manual examination of the bacteria is time-consuming, labor-intensive and costly. Automation of this process will help biologists to get accurate results within a short time. Recently, a lot of work has been happening in biomedical image processing using computer-aided tools and algorithms. Deep learning algorithms are turning out to be a boon in this field of study.  

\par Many researchers in the past have tried to perform classification of bacteria from microscopic images. A lot of data is necessary to perform classification of biomedical images. \par Abdulla et al ~\cite{Abdulla2015}  proposed an algorithm to extract color features from Candida, Gram-Negative Bacilli (GNB) and Gram-Positive Cocci (GPC) bacteria appearing in urine smear images for automatic classification of these objects. They used 60 sample images of 100X resolution (20 each from three bacteria types). Andreini et al~\cite{Andreini2018} created synthetic images from blood agar to classify bacteria. Panicker et al ~\cite{Panicker2018} performed classification of tuberculosis bacteria from sputum smear images.  
\par  Zielinski et al ~\cite{Zielinski2018} created the DIBaS dataset, a Digital Image of Bacterial Species, to perform bacterial colony classification. It consists of 20 images of bacteria belonging to 33 genera and species. Hence, The dataset has 660 images. The authors have made it available for public use and many researchers are using it to perform automatic classification of bacteria from microscopic images. ~\cite{Mohamed2018,Dawid2019} classified bacteria into different shapes, species,etc using this dataset.
\par  Recently, ~\cite{Borowa2021,Iida2020,Shaily2020,Talo2019} detected bacteria using the various ResNet models.  Costa et al~\cite{Costa2020} proposed CNN based deep learning model to segment and identify bacteria type. Dataset was taken from UFAM Pattern Recognition and Optimization Research Group. The images were converted into 9700 patches containing bacteria and 25000 patches without bacteria. The size of each patch was 40X40. 100 such images were converted into 400X400 mosaic images. CNN model was trained using mosaic images and manually annotated labels with a split ratio of 50:25:25. CNN model with Adam optimizer performed best and an F1 score of 98.54\% and accuracy of 98.53\% was recorded. \par
 As seen from the literature, many researchers have proposed CNN models to perform the classification and detection of bacteria but semantic segmentation of the bacteria is not attempted.  Semantic segmentation of bacterial images will aid the biologists to determine the morphological features such as whether the bacteria is gram positive or gram negative, the bacteria are rod-shaped (bacilli) or sphere-shaped(cocci) and are they appearing in clusters or in pairs, etc. However, there is no annotated dataset of bacteria genera and species which is available for open access. Therefore, in this work, we have annotated images from the DIBaS dataset and used it to perform semantic segmentation of the bacteria. This dataset can be used to perform detection and classification of bacteria from microscopic images to diagnose urinary tract infections, tuberculosis, pneumonia, etc. 
\par Standard algorithms which are popularly used for semi-automatic image segmentation are discussed in the next section. After that ResUNet++ architecture for semantic segmentation of microscopic images is explained in section 3. The training procedure and the results are illustrated in section 4 . Conclusion is drawn in section 5.
\section{Semi-Automated Labelling}
Deep learning models need large datasets and their corresponding annotated labels for training. Annotations have to be carried out by experienced biologists. Segmentation of bacteria at pixel level will help the biologists to get deep understanding of their arrangement, shape and stain-type. Also, this will be helpful to recognize the bacteria genera and species.  In this section, annotations of images using k-means clustering and Otsu thresholding algorithms is discussed. Also the importance of morphological operations on the annotated labels is highlighted.
\subsection{k-means clustering}
k-means clustering is an unsupervised segmentation algorithm to segment the dataset into k clusters. It tries to group data points with similar properties into a single cluster. Centroids with the least squared distance between the data points are computed. The data points are assigned to the closest centroid, and data points assigned to the same centroid are treated as a single cluster. For an RGB image, the k-means clustering algorithm is used with RGB pixel values as data points, and the resulting labels of each pixel can be used for image segmentation. \par
Due to staining, bacteria have a significantly different color than the background. Stain is applied uniformly for all the bacteria, which makes the color of bacteria uniform throughout the image. The fact that bacteria color is both distinct and uniform can be exploited because pixels of bacteria form a cluster in a 3D RGB space. The k-means clustering algorithm can label the cluster to segment bacteria in the images.
 \begin{figure}[!h]
    \centering
    \subfloat[Before K-means]{\frame{\includegraphics[width=5cm,height=5cm]{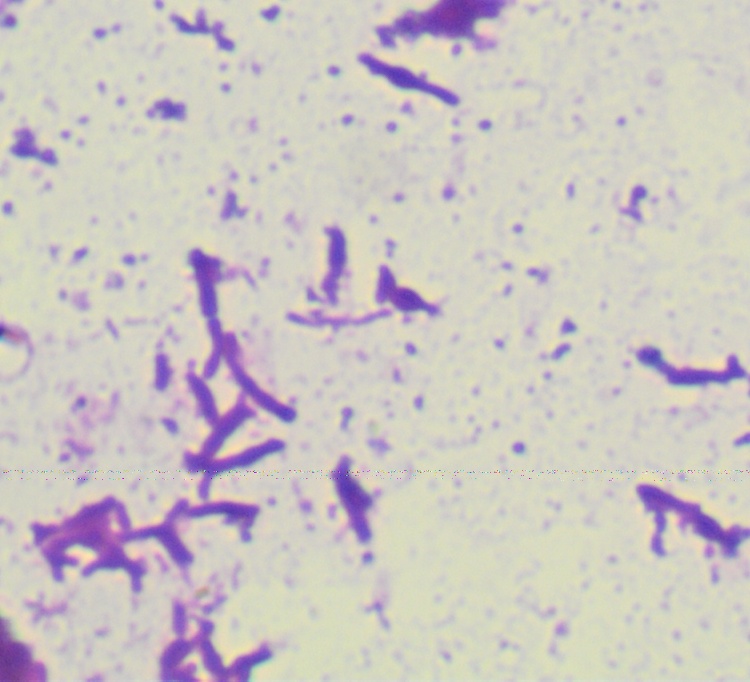} }}
     \qquad
    \subfloat[After K-means]{\frame{\includegraphics[width=5cm,height=5cm]{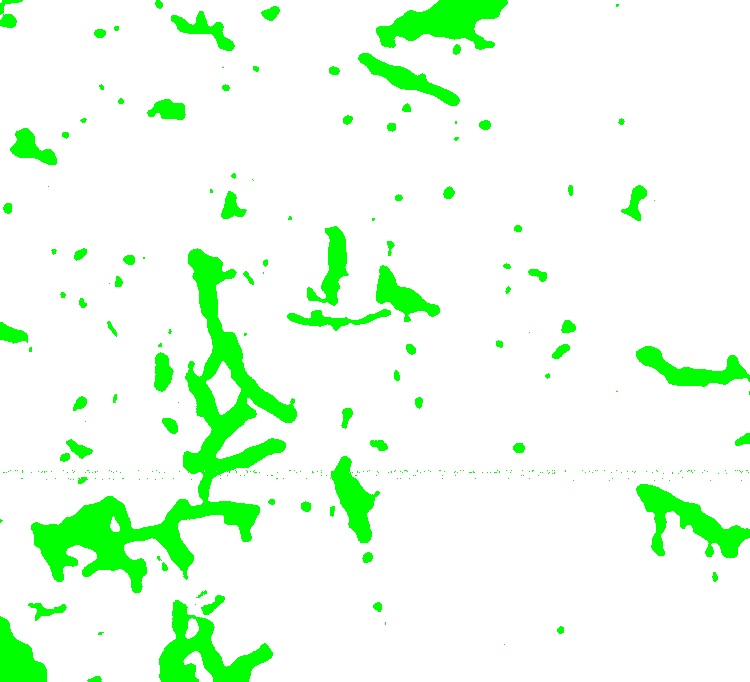} }}
    \caption{Performing k-means on microscopic image dataset}
    \label{fig:kmeans}
    
\end{figure}
\subsection{Otsu thresholding}
Otsu algorithm is widely used to perform image segmentation. It generates a threshold value based on a weighted variance of the foreground and background classes, and this threshold helps distinguish the foreground pixels from the background class.
\par
The k-means algorithm works when bacteria have a uniform color in an image, but it is not the case for the pictures of veionella bacteria in the DIBaS dataset. The color of the stain varies significantly throughout the image; hence otsu thresholding is used to segment the bacteria from the image.

 \begin{figure}[h]
    \centering
    \subfloat[Before Otsu thresholding]{\frame{\includegraphics[width=5cm,height=5cm]{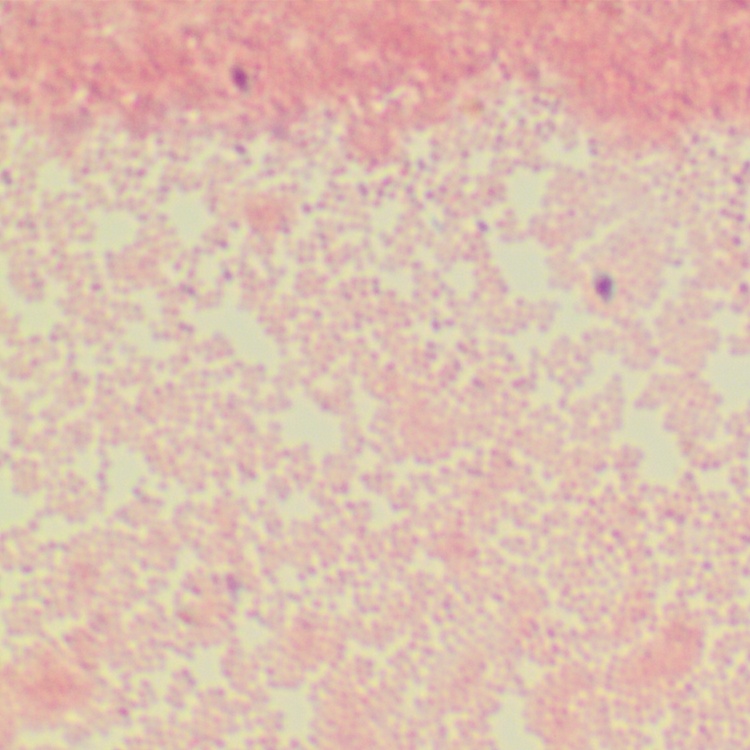} }}
     \qquad
    \subfloat[After Otsu thresholding]{\frame{\includegraphics[width=5cm,height=5cm]{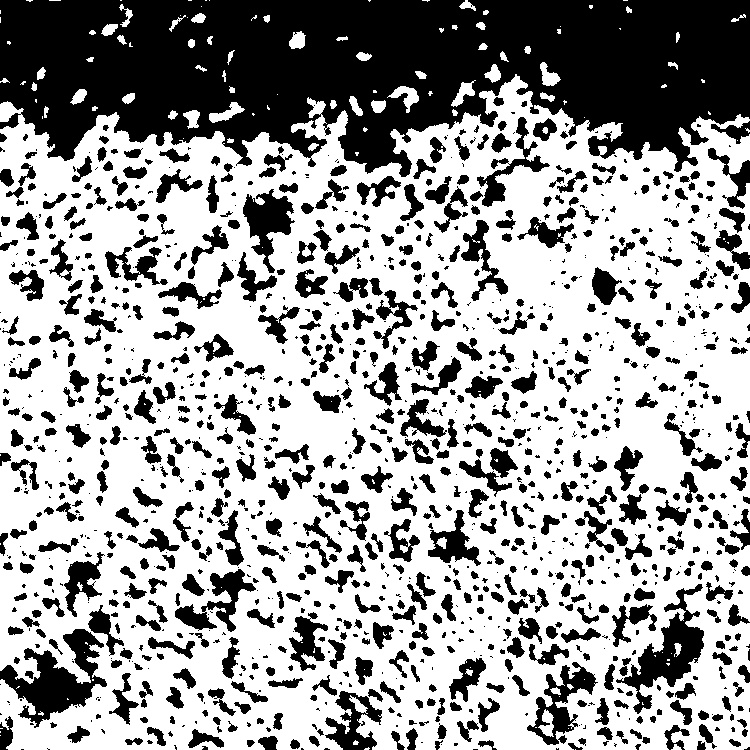} }}
    \caption{Performing otsu thresholding on a microscopic image}
    \label{fig:otsu}
\end{figure}

\subsection{Morphological operations}
Morphological closing can be formed by 2 basic mathematical morphology operations dilation and erosion. Morphological operations help to remove artifacts and noise from the images. \par
 \begin{figure}[h]
    \centering
    \subfloat[Before morphological closing]{\frame{\includegraphics[width=5cm,height=5cm]{k2m9msk.png} }}
     \qquad
    \subfloat[After morphological closing]{\frame{\includegraphics[width=5cm,height=5cm]{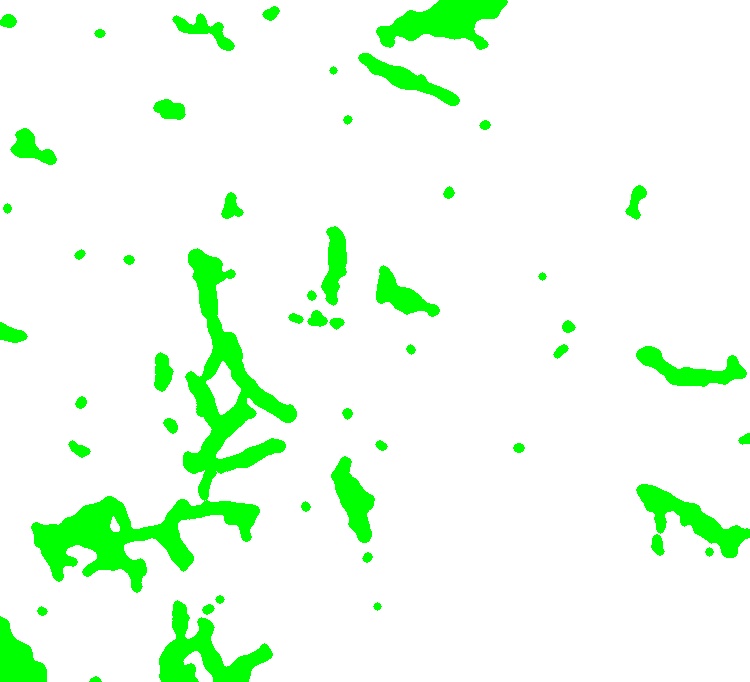} }}
    \caption{Performing morphological closing operation to the image on left has removed staining artifacts as seen in the right image}
    \label{fig:morph}
\end{figure}
The stains in the bacterial images are the same color as bacteria; hence, the stain artifacts get clustered with bacteria in the K-means algorithm. But they are significantly smaller when compared to bacteria, so morphological closing can be used to remove the stain artifacts. This is illustrated in figure \ref{fig:morph}. The image on the left has some small patches of noisy regions. When a morphological closing operation is applied on this image using circular kernel, these artifacts are reduced as seen in the right image.

\section{Semantic Segmentation of Bacterial Images using ResUNet++}
\par Automatic segmentation of bacteria is quick, free of human intervention and also it gives accurate results. For instance, automatic detection of bacteria and pus cells from gram-stained urine smear images will aid the microbiologists in the diagnosis of urinary tract infection. In the absence of these particles in smear, the patient need not undergo urine culture test. Since culture test results need atleast 48 hours, this will reduce the time, cost as well as the labour involved in culture examination and testing. Also the results are available in less time. Therefore, we perform automatic semantic segmentation of bacteria from DIBaS dataset. \par
U-Net~\cite{ronneberger2015unet} is a popular deep learning architecture for semantic segmentation of biomedical data.
The ResUNet++~\cite{KHANNA20201314,Siddique_2021} architecture is a slightly modified version of U-Net and it provides state-of-the-art results.  ResUNet++ architecture ~\cite{ResUNetPP} contains one stem block followed by three encoder blocks, ASPP, and three decoder blocks. This architecture takes advantage of the residual blocks, the squeeze and the excitation block, Atrous Spatial Pyramidal Pooling (ASPP) and the attention block. The Residual block helps to build a deeper neural network which solves the degradation problem in all the encoder units. Batch normalization unit, A Rectified Linear Unit (ReLU) activation unit and convolutional layers are present in every residual block.  This model provides accurate results in less time when compared to other variants of ResUNet++. Therefore, we choose ResUNet++ model for segmenting bacteria from DIBaS dataset.

\section{Evaluation metrics}
Accuracy, F1 score, precision, recall and Intersection-over-Union (IoU) are the frequently used metrics to evaluate the model performance over the test dataset. A brief introduction to these is provided in this section. In a binary classification problem, there will be two classes, positive and negative class. The model can predict the labels are either one of these two classes. Following are the commonly used terms while defining the performance metrics.
\begin{itemize}
    \item \textbf{True Positive (TP)}: The predicted label and the actual label both belong to the positive class.
    \item \textbf{True Negative (TN)}: The predicted label and the actual label both belong to the actual negative class.
    \item \textbf{False Positive (FP)}: The predicted label is positive class but the actual label is negative class.
    \item \textbf{False Negative (FN)}: The predicted label is negative class but the actual label is positive class.
\end{itemize}
\subsection*{Accuracy:}
Accuracy gives the measure of the model performance over all the classes. This metric is useful when we deal with the balanced dataset. 

\begin{equation}
\label{eq_acc}
Accuracy = \frac{TP+TN}{TP+FP+FN+TN}
\end{equation}

Accuracy is best if the value is 100 percentage and worst if it is 0 percentage. If the dataset used in training the model is imbalanced then even if the performance of one of the classes is poor the overall accuracy will be high. Therefore most of the time this metric is not used as a standalone performance evaluation parameter.
\subsection*{Precision:}
This metric tells, out of all the predictions made as positive by the model how many are actually positive. 
\begin{equation}
\label{eq_prec}
Precision = \frac{TP}{TP+FP}
\end{equation}
Precision value can be between 0 and 1, 0 being worst case and 1 indicates perfect precision. Higher precision value means that the model has made high correct positive predictions or very low incorrect positive predictions.
\subsection*{Recall:}
Recall gives the information about out of the total positive cases, how many are correctly predicted by the model. 
\begin{equation}
\label{eq_rec}
Recall = \frac{TP}{TP+FN}
\end{equation}
Recall value can be between 0 and 1, 0 being worst recall case and 1 indicates perfect recall. The model gives recall of 100 \% when all positive labels are correctly classified even if all the negative labels are misclassified.
\subsection*{F1 Score:}
F1 score is defined as the harmonic mean of precision and recall.
\begin{equation}
\label{eq_f1}
F1~score = \frac{2*precision*recall}{precision+recall}
\end{equation}
F1 score has value between 0 and 1. F1 score of 1 is best and the worst value is 0. This metric is reliable when dealing with imbalanced dataset.

\subsection*{Intersection-over-Union(IoU) Score:} IoU is the ratio of the area of overlap between groundtruth and predicted image to the area of the union between these two images. 

\begin{equation}
\label{eq_iou}
IoU = \frac{TP}{TP+FP+FN}
\end{equation}

IoU score of 1 is best and 0 is worst.

\section{Experiments and Results}
The process involves the development of labels of images from DIBaS dataset. This is a semi-automatic labelling based on k-means clustering, otsu thresholding and morphological closing operations. The images with the generated labels are used to obtain semantic segmentation of bacterial images. ResUNet++ model is used and this is a completely automatic process.
\subsection{Dataset}
Gram-stained images of different bacterial genera and species have been taken from Digital Images of Bacteria Species(DIBaS) dataset~\cite{Zielinski2018}. The dataset consists of 33 genera and species of bacteria with each species having 20 images. So there are 660 images of size 2048X1532 and 100X magnification and they are available for public access in \href{https://github.com/gallardorafael/DIBaS-Dataset}{this link}. All the images are annotated using k-means clustering and otsu thresholding algorithms.

\subsection{Semi-automatic Annotation}
 
Some bacteria species have a lot of stain artifacts. Using morphological closing to remove those artifacts is not possible because the size of stains is sometimes more than bacteria, so to remove those artifacts, k means clustering with k=3 is used in the hope of artifacts getting clustered in 3rd cluster. But this can't be done for all the bacteria species as some bacteria species have a nonuniform color of staining; hence some bacteria get involved in artifact cluster. To overcome this problem, all the bacteria species were examined manually to check whether k=3 or k=2 gave a better result. 
\begin{figure}[]
  \centering
  \setlength{\tabcolsep}{12pt}
  \begin{tabular}{ccc}
   
    \frame{\includegraphics[width=1.5in]{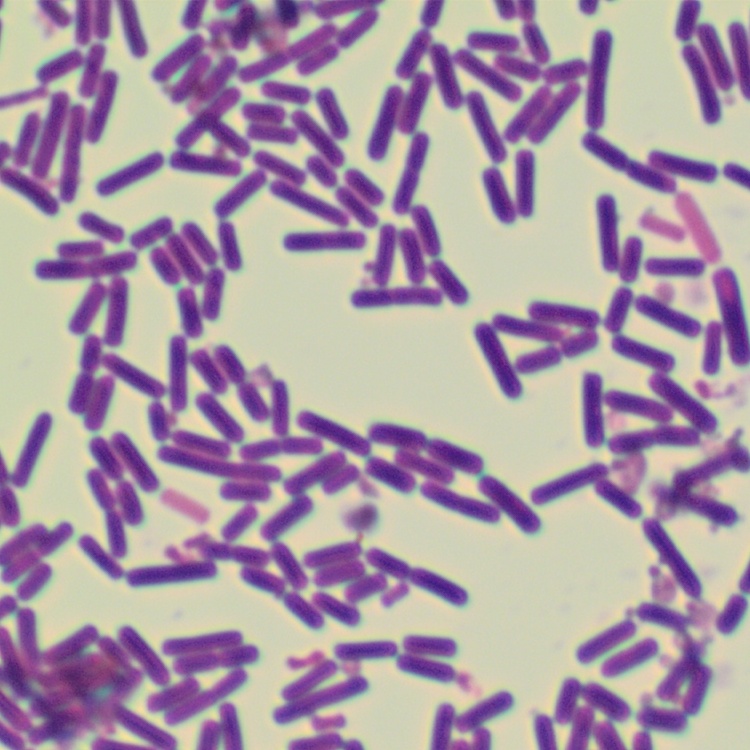}} \hfill \hfill &

    \frame{\includegraphics[width=1.5in]{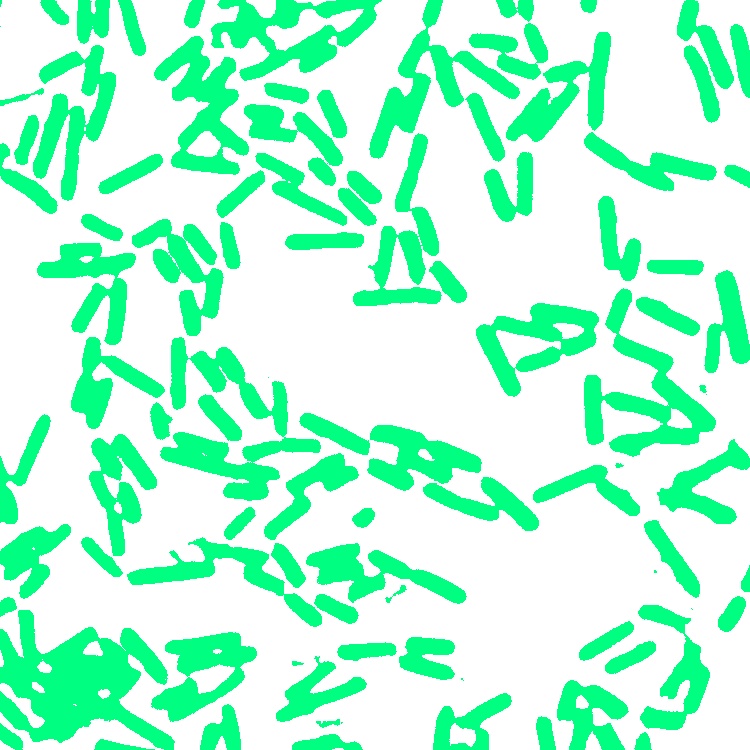}} \hfill \hfill 
    &\frame{\includegraphics[width=1.5in]{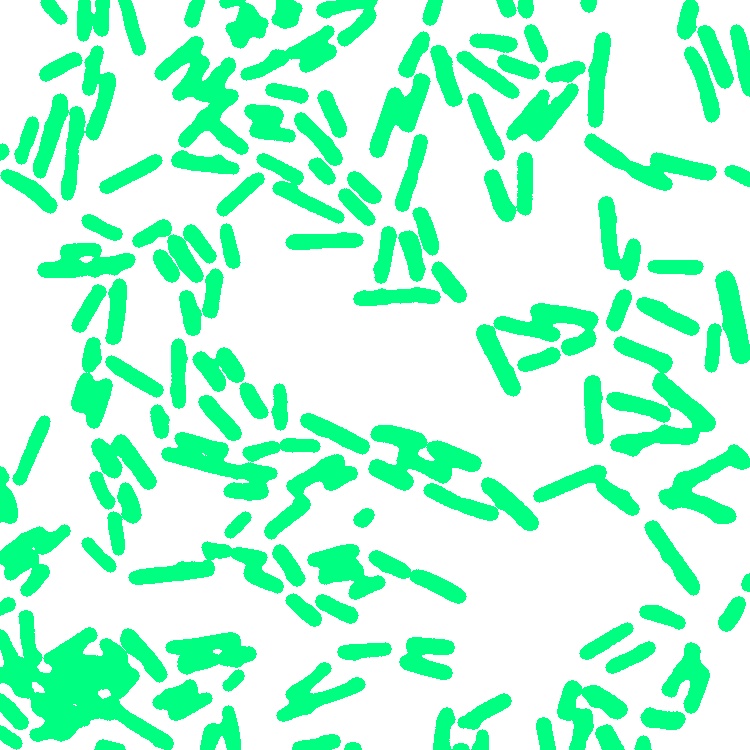}}\\
    
    \textbf{Clostridium.perfringens} & \textbf{kmeans with k=3} & \textbf{kernel size of 13X13} \\

    \frame{\includegraphics[width=1.5in]{k2m9img.png}} \hfill \hfill &

    \frame{\includegraphics[width=1.5in]{k2m9msk.png}} \hfill \hfill
    &\frame{\includegraphics[width=1.5in]{k2m9mskmorph.png}}
    \\
    
    \textbf{Bifidobacterium.spp} & \textbf{kmeans with k=2} & \textbf{kernel size of 9X9} \\

    \frame{\includegraphics[width=1.5in]{otsuimg.png}} \hfill \hfill&

    \frame{\includegraphics[width=1.5in]{otsumsk.png}} \hfill \hfill
    &\frame{\includegraphics[width=1.5in]{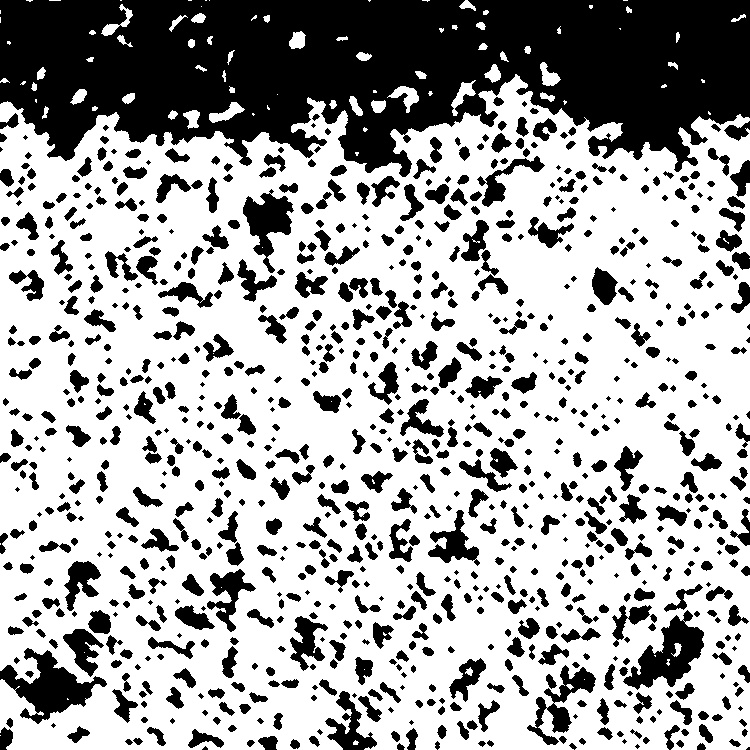}}
    \\
    
    \textbf{Veionella} & \textbf{otsu thresholding} & \textbf{kernel size of 5X5} \\

    \\
    \textbf{(a)Bacteria Image} & \textbf{(b)Image segmentation} & \textbf{(c)Morphological closing}
  \end{tabular}
 \caption{\textbf{Sample images with their ground truth labels}}
 \label{annotated_labels}
\end{figure}
Bacteria species have very different sizes, so a single kernel size can't be chosen for all the species to perform morphological closing. To overcome this problem best kernel size specific to bacteria species was selected by manually examining various kernel sizes.Few sample images are shown in Figure \ref{annotated_labels}. The annotated labels can be accessed from \href{https://github.com/nitk-smile/DIBaS_Annotations}{https://github.com/nitk\-smile/DIBaS\_Annotations}.

\subsection{Training using ResUNet++}
\begin{figure}[]
    \centering
    \subfloat[loss]{\frame{\includegraphics[width=8cm]{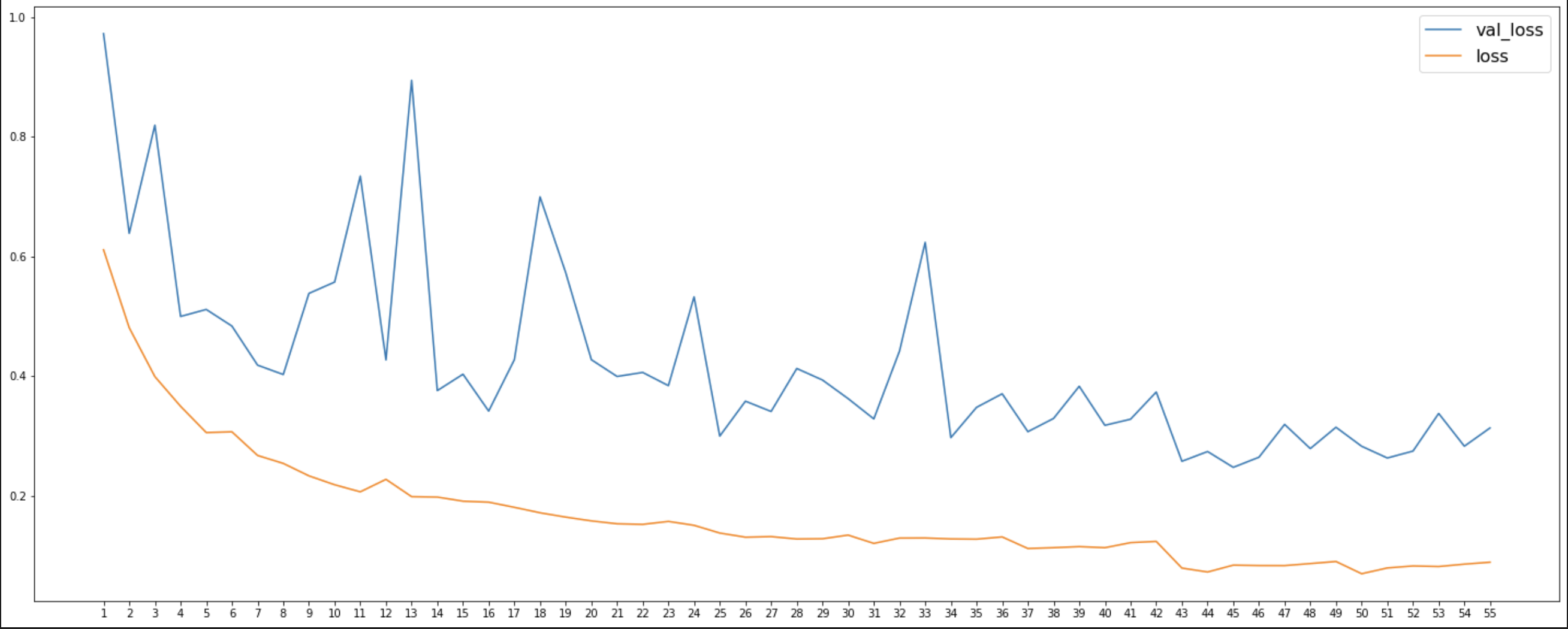} }}
     \qquad
    \subfloat[categorical accuracy]{\frame{\includegraphics[width=8cm]{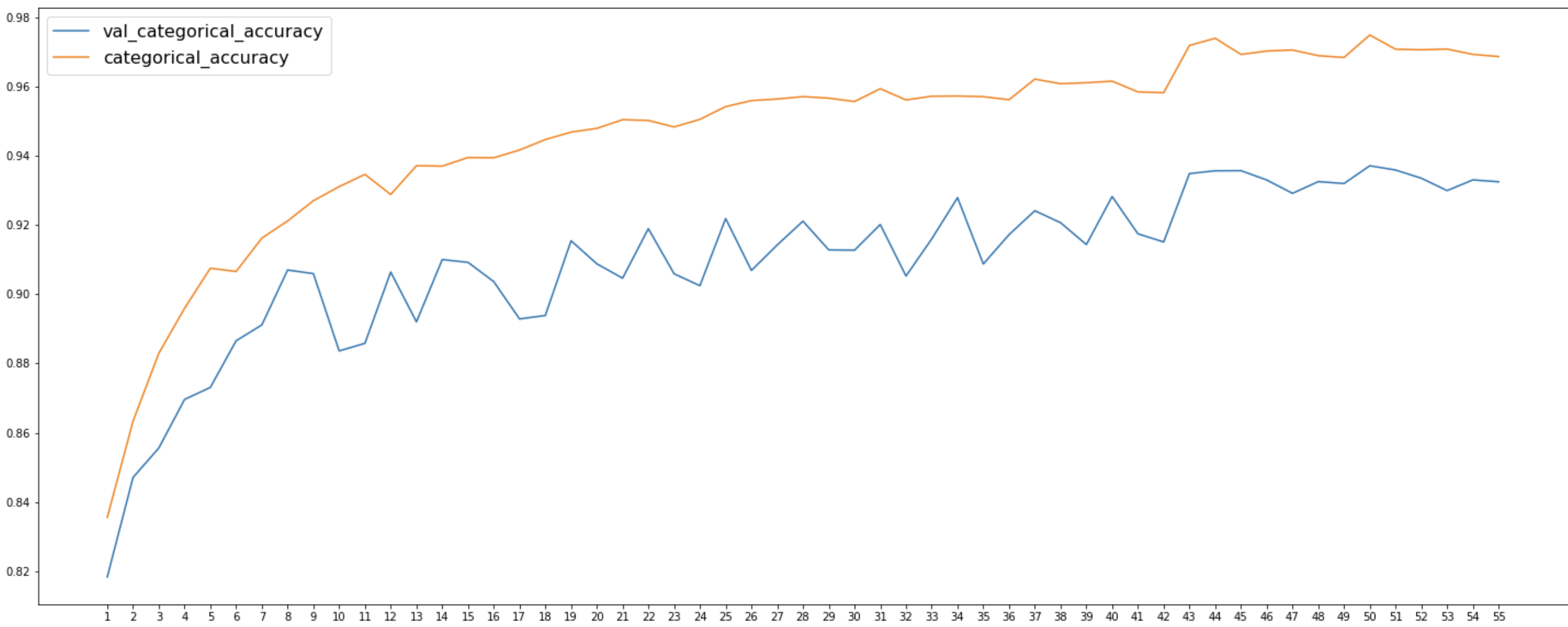} }}
    \caption{Metrics over 54 epochs}
    \label{fig:loss}
    
\end{figure}

The ResUNet++ model is implemented on Tensorflow's Keras platform. The training is performed by splitting the dataset into train:validation:test as 60:20:20. The images in the training dataset are converted into overlapping patches while those in validation set are converted to non-overlapping patches of size 512X512. Data augmentation is performed using rotation, random flipping, and shifting. A batch size of 4 is chosen. Categorical accuracy and F1 score are used as performance metrics. Categorical cross-entropy is used as a loss function. RMS prop optimizer is used, with the learning rate set to 0.0001 for 40 epochs and decreased to 0.00001 for later epochs. The model was trained for 50 epochs. The loss and training accuracy over 54 epoch is depicted in Figure \ref{fig:loss}.
The training was stopped after 50 epochs as the loss started to increase after that and the change in loss was minimal after 34 epochs. genera and species-wise accuracy, F1 score and IoU scores are tabulated in Table \ref{table_results}.
\

\begin{table}[]
\caption{Bacteria genera-wise and species-wise performance scores}
\label{table_results}
\begin{tabular}{|c|c|c|c|c|c|c|}
\hline
\textbf{Sl. No.} & \textbf{bacteria}            & \textbf{precision} & \textbf{recall} & \textbf{fscore} & \textbf{IoU}  & \textbf{Accuracy} \\\hline
1                & Acinetobacter baumanii       & 0.97               & 0.69            & 0.81            & 0.68          & 0.97              \\
2                & Actinomyces israeli          & 1.00               & 0.96            & 0.98            & 0.95          & 0.97              \\
3                & Bacteroides fragilis         & 0.93               & 0.67            & 0.78            & 0.64          & 0.97              \\
4                & Bifidobacterium spp          & 0.70               & 0.98            & 0.81            & 0.69          & 0.96              \\
5                & Candida albicans             & 0.98               & 0.94            & 0.96            & 0.92          & 0.96              \\
6                & Clostridium perfringens      & 0.95               & 0.96            & 0.95            & 0.91          & 0.98              \\
7                & Enterococcus faecium         & 0.79               & 0.55            & 0.65            & 0.48          & 0.97              \\
8                & Enterococcus faecium         & 0.96               & 0.87            & 0.91            & 0.84          & 0.97              \\
9                & Escherichia coli             & 0.09               & 0.98            & 0.17            & 0.09          & 0.96              \\
10               & Fusobacterium                & 0.81               & 0.99            & 0.89            & 0.80          & 0.96              \\
11               & Lactobacillus casei          & 0.90               & 0.82            & 0.86            & 0.75          & 0.97              \\
12               & Lactobacillus crispatus      & 0.90               & 0.99            & 0.94            & 0.89          & 0.97              \\
13               & Lactobacillus delbrueckii    & 0.75               & 0.49            & 0.59            & 0.42          & 0.78              \\
14               & Lactobacillus gasseri        & 0.88               & 0.98            & 0.93            & 0.87          & 0.98              \\
15               & Lactobacillus jehnsenii      & 0.99               & 0.44            & 0.61            & 0.44          & 0.95              \\
16               & Lactobacillus johnsonii      & 0.18               & 0.98            & 0.30            & 0.18          & 0.97              \\
17               & Lactobacillus paracasei      & 0.56               & 0.95            & 0.71            & 0.55          & 0.97              \\
18               & Lactobacillus plantarum      & 0.86               & 0.78            & 0.82            & 0.70          & 0.97              \\
19               & Lactobacillus reuteri        & 0.00               & 0.12            & 0.01            & 0.00          & 0.97              \\
20               & Lactobacillus rhamnosus      & 0.87               & 0.35            & 0.49            & 0.33          & 0.97              \\
21               & Lactobacillus salivarius     & 1.00               & 0.87            & 0.93            & 0.87          & 0.97              \\
22               & Listeria monocytogenes       & 0.98               & 0.99            & 0.99            & 0.98          & 0.97              \\
23               & Micrococcus spp              & 0.98               & 0.97            & 0.98            & 0.95          & 0.97              \\
24               & Neisseria gonorrhoeae        & 0.87               & 0.94            & 0.90            & 0.82          & 0.94              \\
25               & Porfyromonas gingivalis      & 0.43               & 0.39            & 0.41            & 0.26          & 0.97              \\
26               & Propionibacterium acnes      & 0.95               & 0.86            & 0.90            & 0.82          & 0.98              \\
27               & Proteus                      & 0.97               & 0.73            & 0.83            & 0.71          & 0.97              \\
28               & Pseudomonas aeruginosa       & 0.74               & 0.94            & 0.83            & 0.71          & 0.98              \\
29               & Staphylococcus aureus        & 0.57               & 0.77            & 0.65            & 0.48          & 0.88              \\
30               & Staphylococcus epidermidis   & 0.36               & 0.37            & 0.37            & 0.23          & 0.86              \\
31               & Staphylococcus saprophiticus & 0.60               & 0.95            & 0.74            & 0.58          & 0.87              \\
32               & Streptococcus agalactiae     & 0.99               & 0.93            & 0.96            & 0.92          & 0.97              \\
33               & Veionella                    & 0.98               & 0.77            & 0.87            & 0.76          & 0.87              \\ \hline
                 & \textbf{Average Scores}      & \textbf{0.74}      & \textbf{0.79}   & \textbf{0.77}   & \textbf{0.64} & \textbf{0.95}    \\ \hline
\end{tabular}
\end{table}

\par
\begin{figure}
  \centering
  \setlength{\tabcolsep}{8pt}
  \begin{tabular}{cccc}
   \textbf{(a)Bacteria Image} & \textbf{(b)Groundtruth} & \textbf{(c)Predicted Label} &\textbf{(d)Predicted class}\\
   \\
    \frame{\includegraphics[width=1.2in]{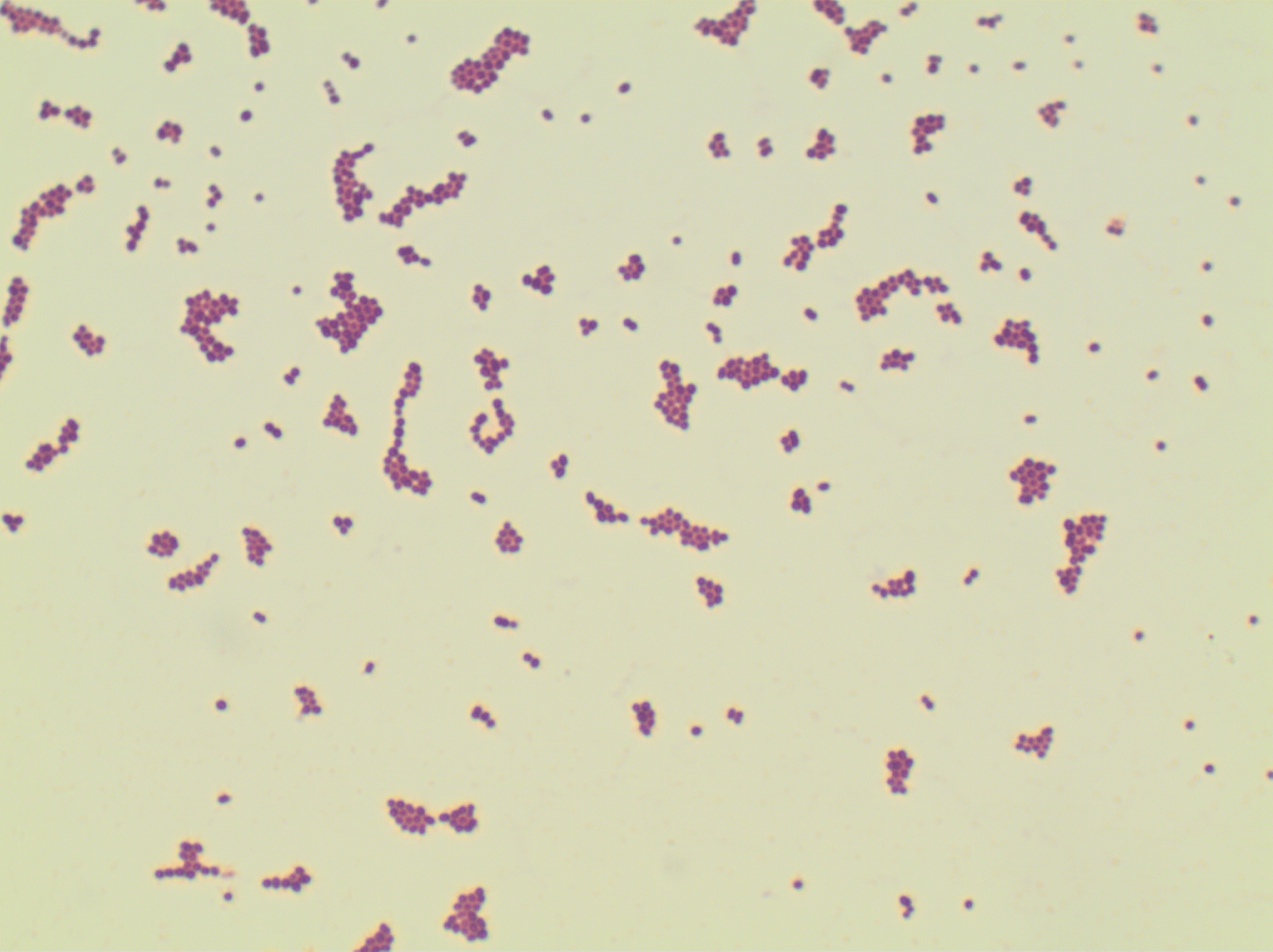}} \hfill \hfill &

    \frame{\includegraphics[width=1.2in]{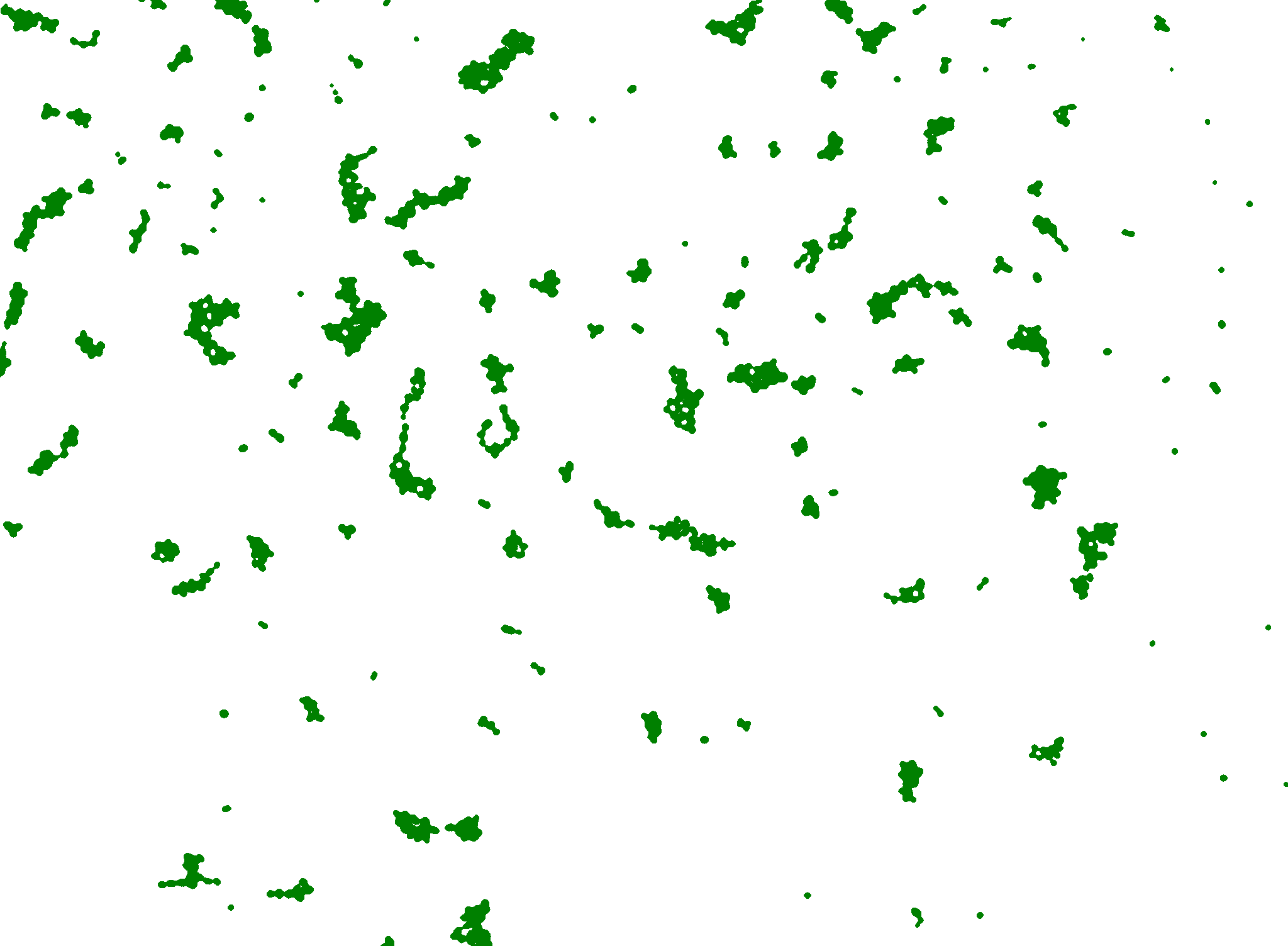}} \hfill \hfill 
    &\frame{\includegraphics[width=1.2in]{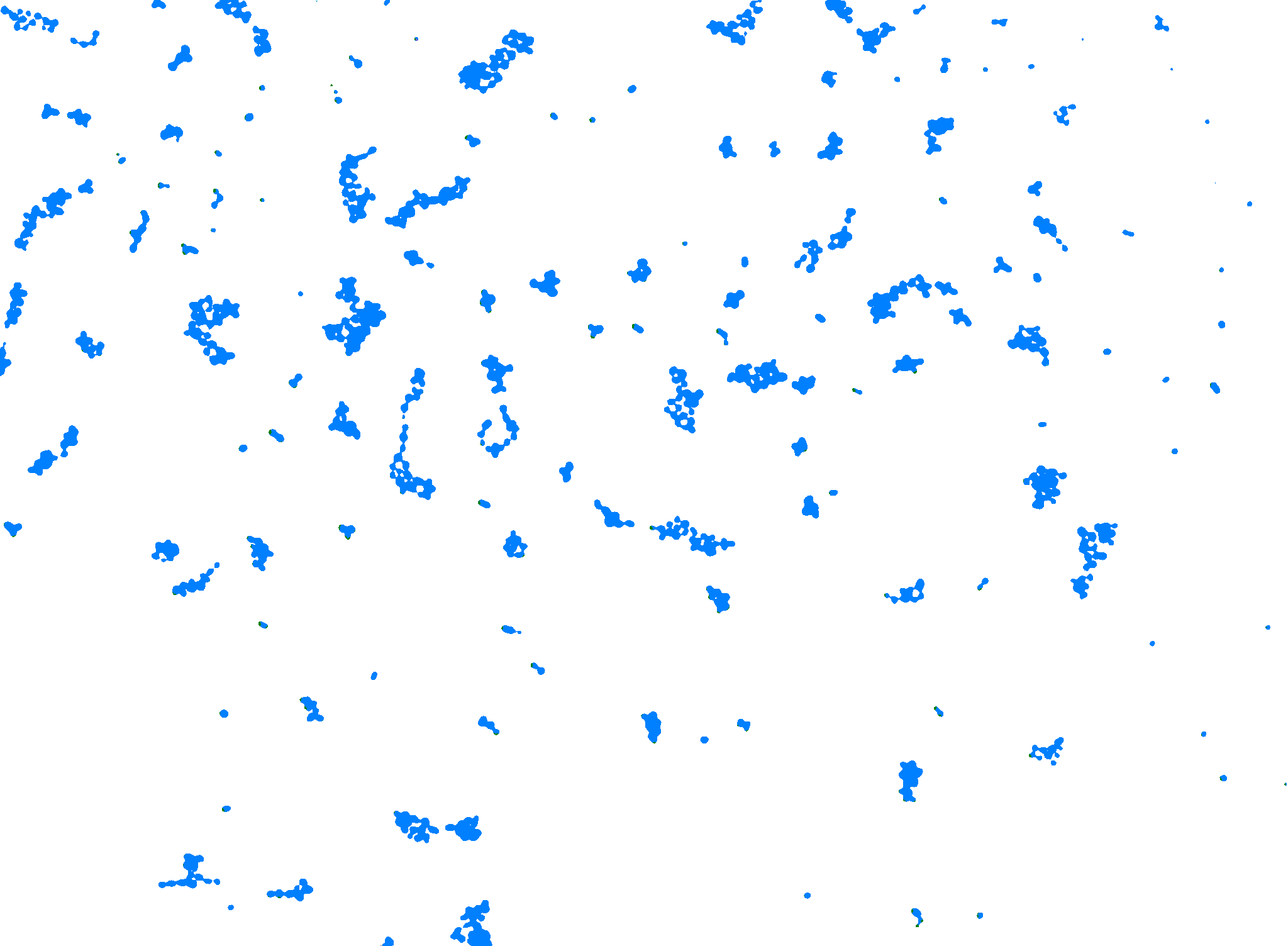}} &\frame{\includegraphics[width=1.2in]{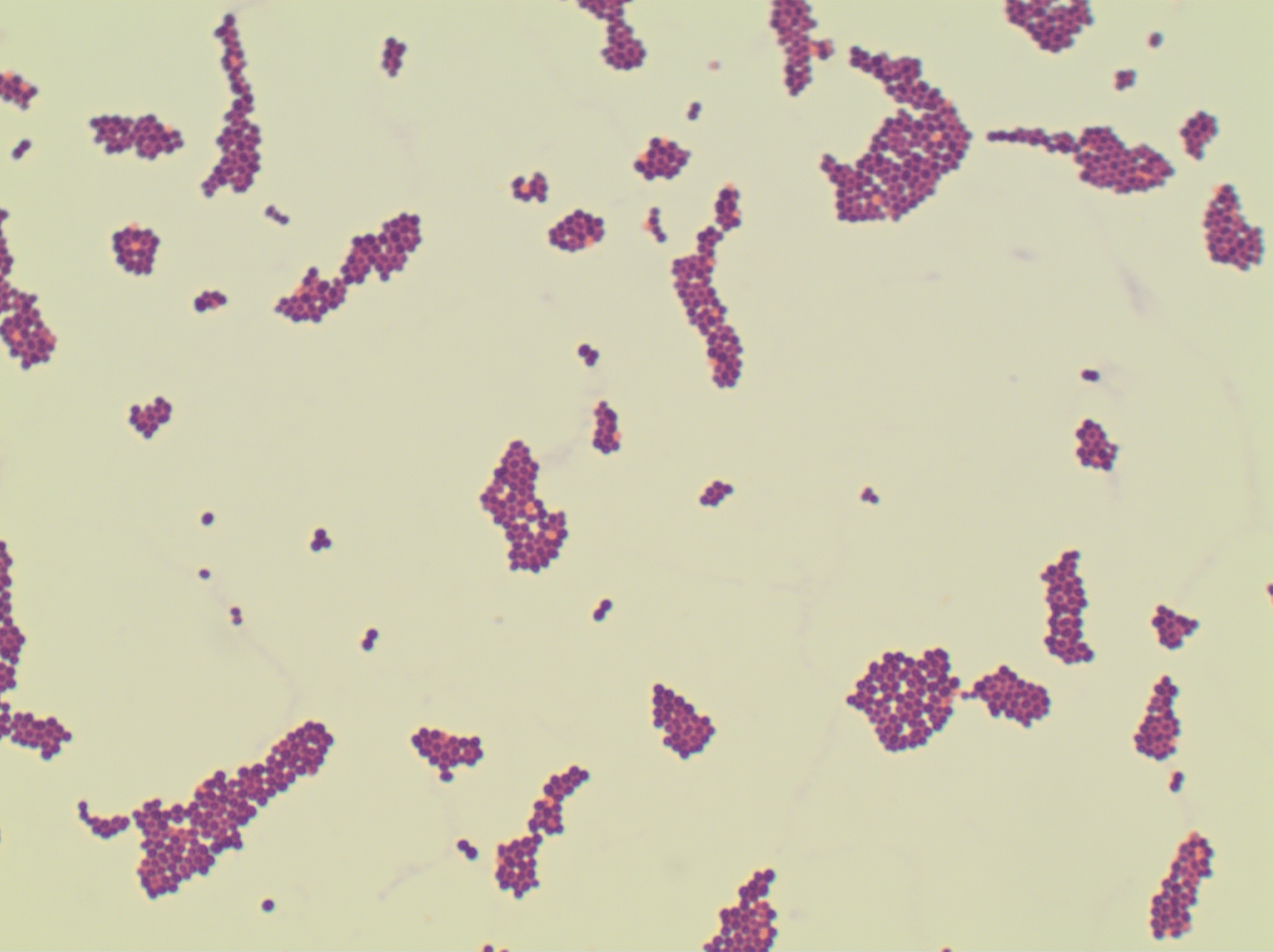}}\\
    
    \textbf{Enterococcus} &  \textbf{Enterococcus} &  \textbf{Enterococcus} & \textbf{Enterococcus}\\
    
    \textbf{faecalis} &  \textbf{faecalis gt} &  \textbf{faecium label} & \textbf{faecium}\\

    \frame{\includegraphics[width=1.2in]{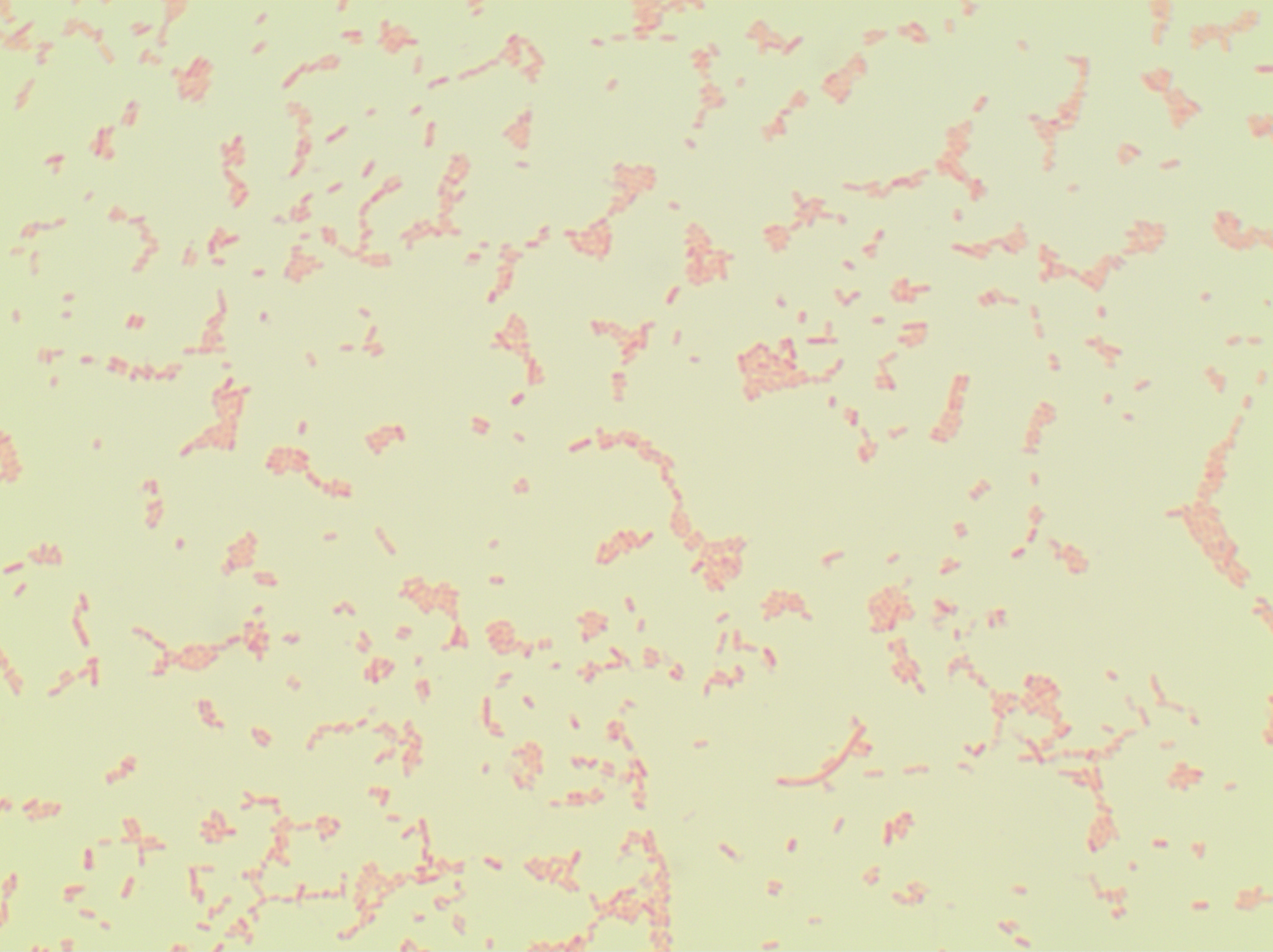}} \hfill \hfill &

    \frame{\includegraphics[width=1.2in]{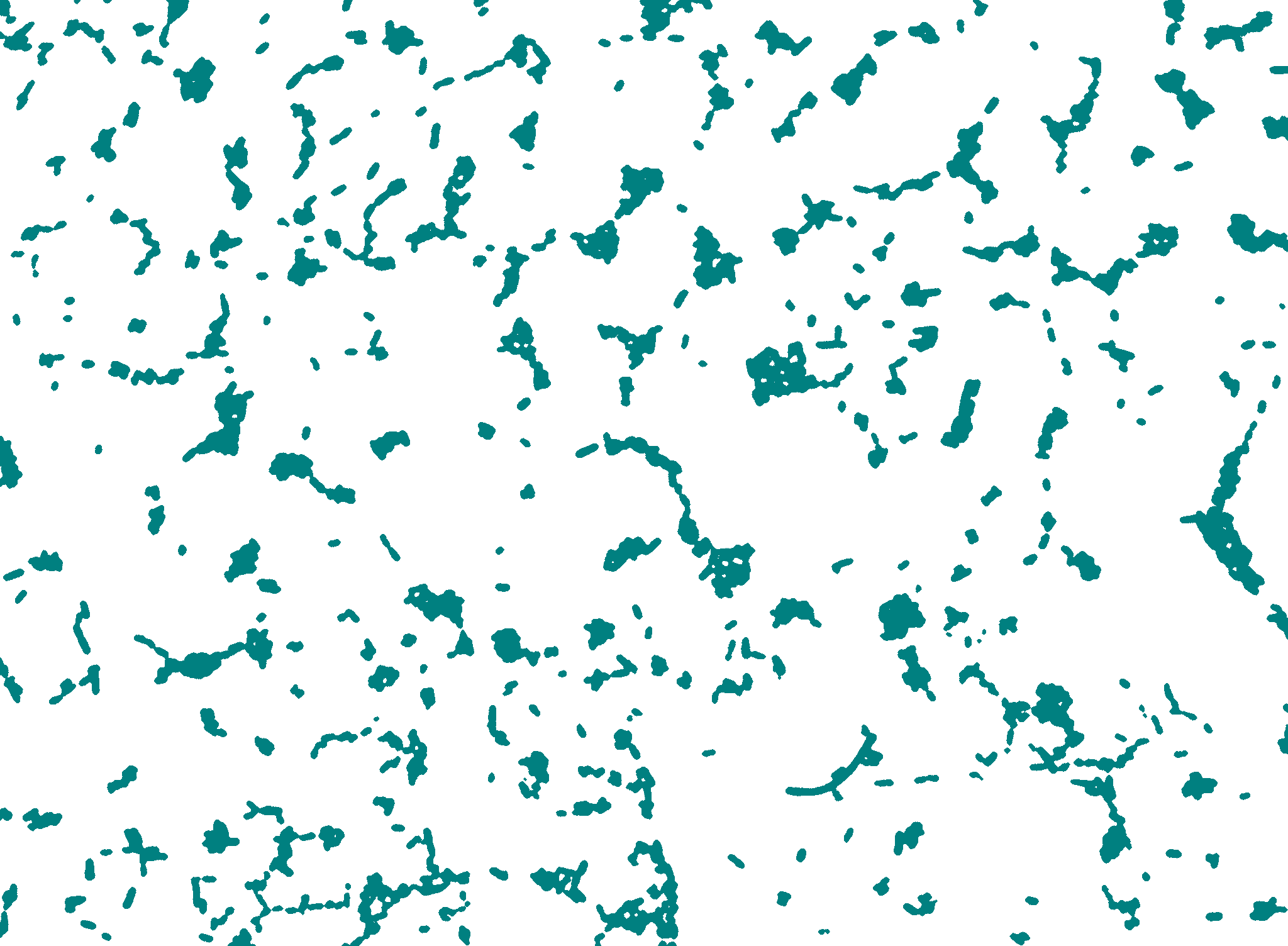}} \hfill \hfill
    &\frame{\includegraphics[width=1.2in]{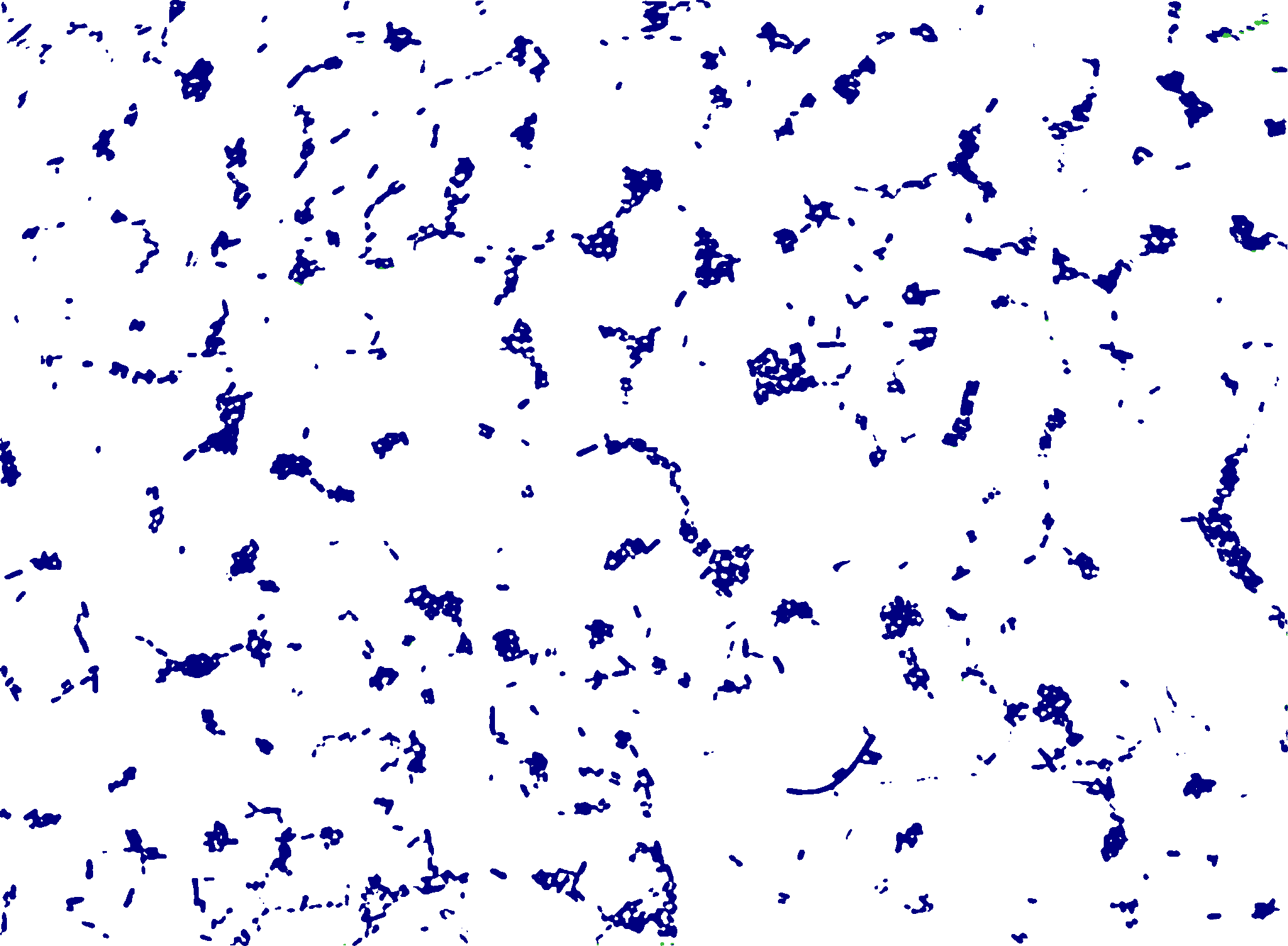}}
    &\frame{\includegraphics[width=1.2in]{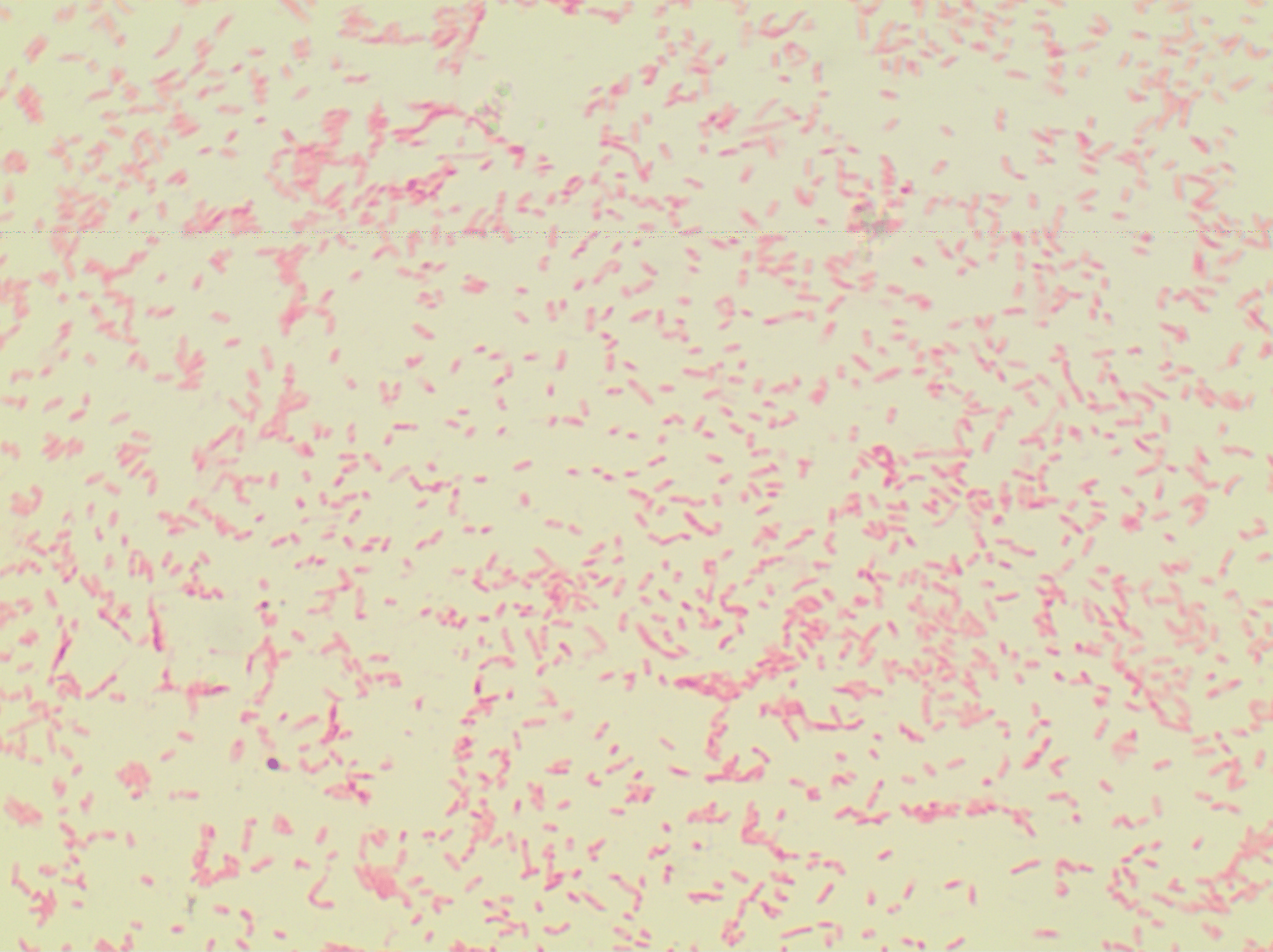}}
    \\
    
    \textbf{Escherichia} & \textbf{Escherichia } & \textbf{Bacteroides}
    &\textbf{Bacteroides}\\
    
    \textbf{coli} & \textbf{coli gt} & \textbf{fragilis label}
    &\textbf{fragilis}\\

    \frame{\includegraphics[width=1.2in]{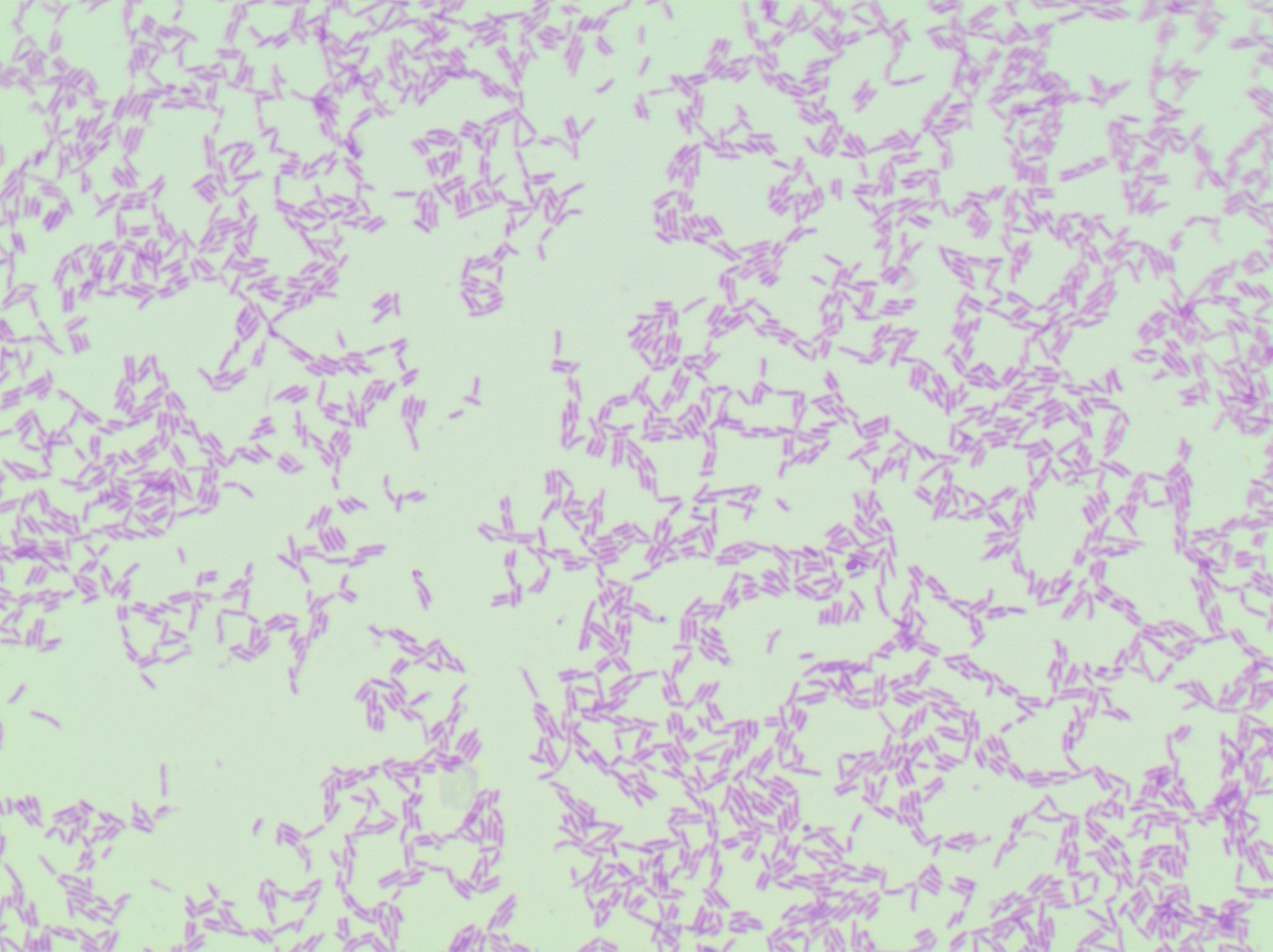}} \hfill \hfill&

    \frame{\includegraphics[width=1.2in]{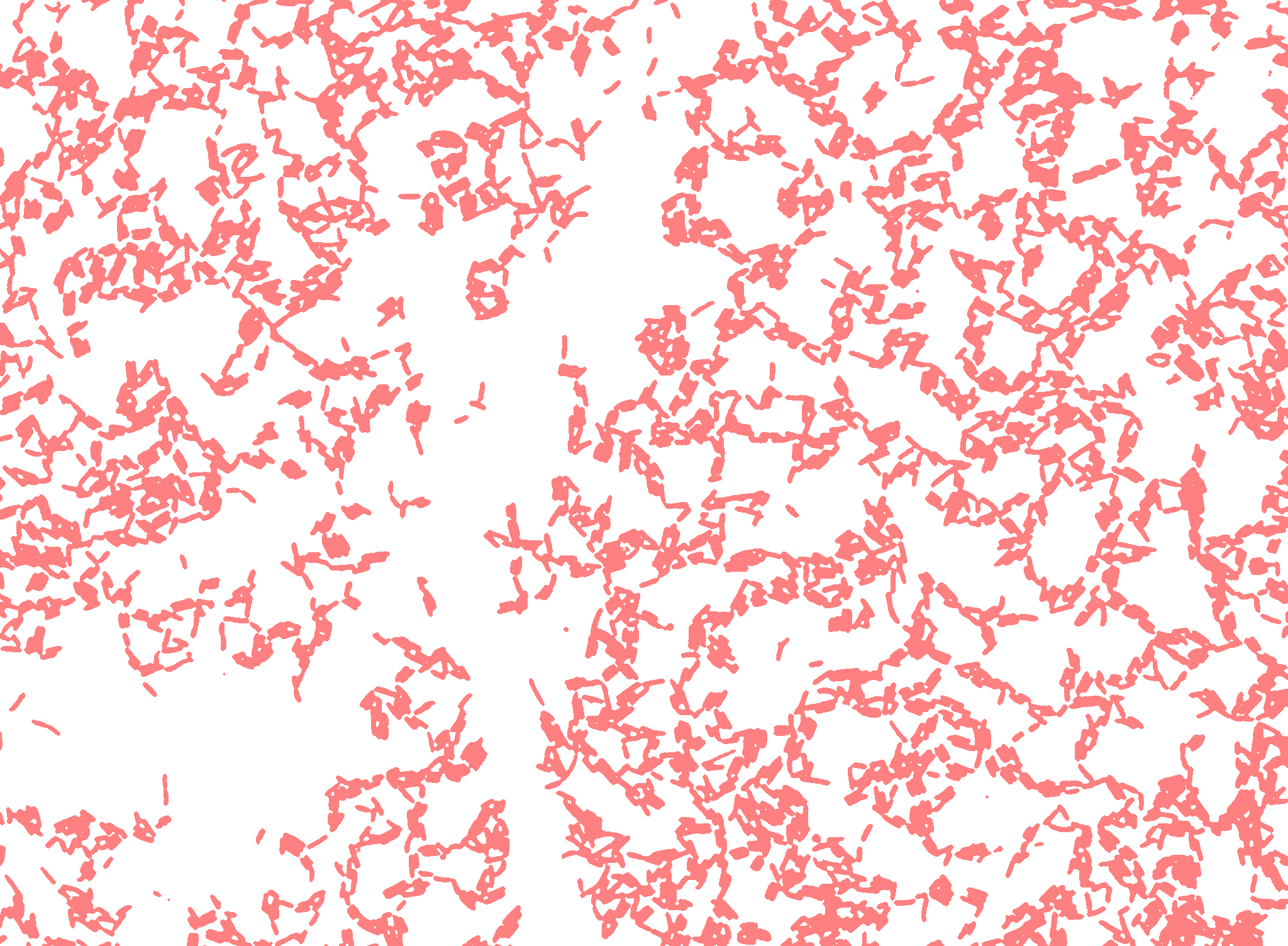}} \hfill \hfill
    &\frame{\includegraphics[width=1.2in]{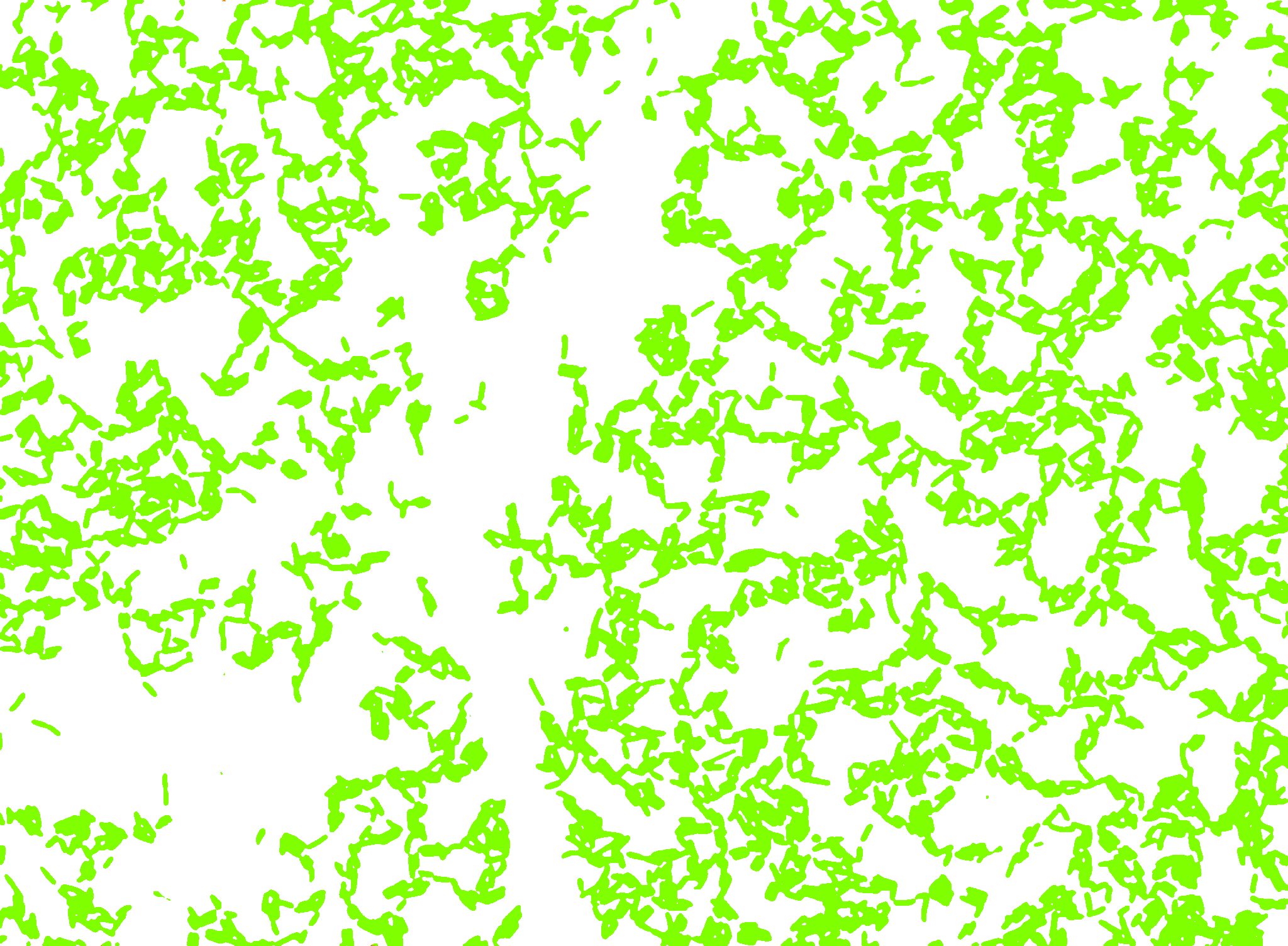}}
    &\frame{\includegraphics[width=1.2in]{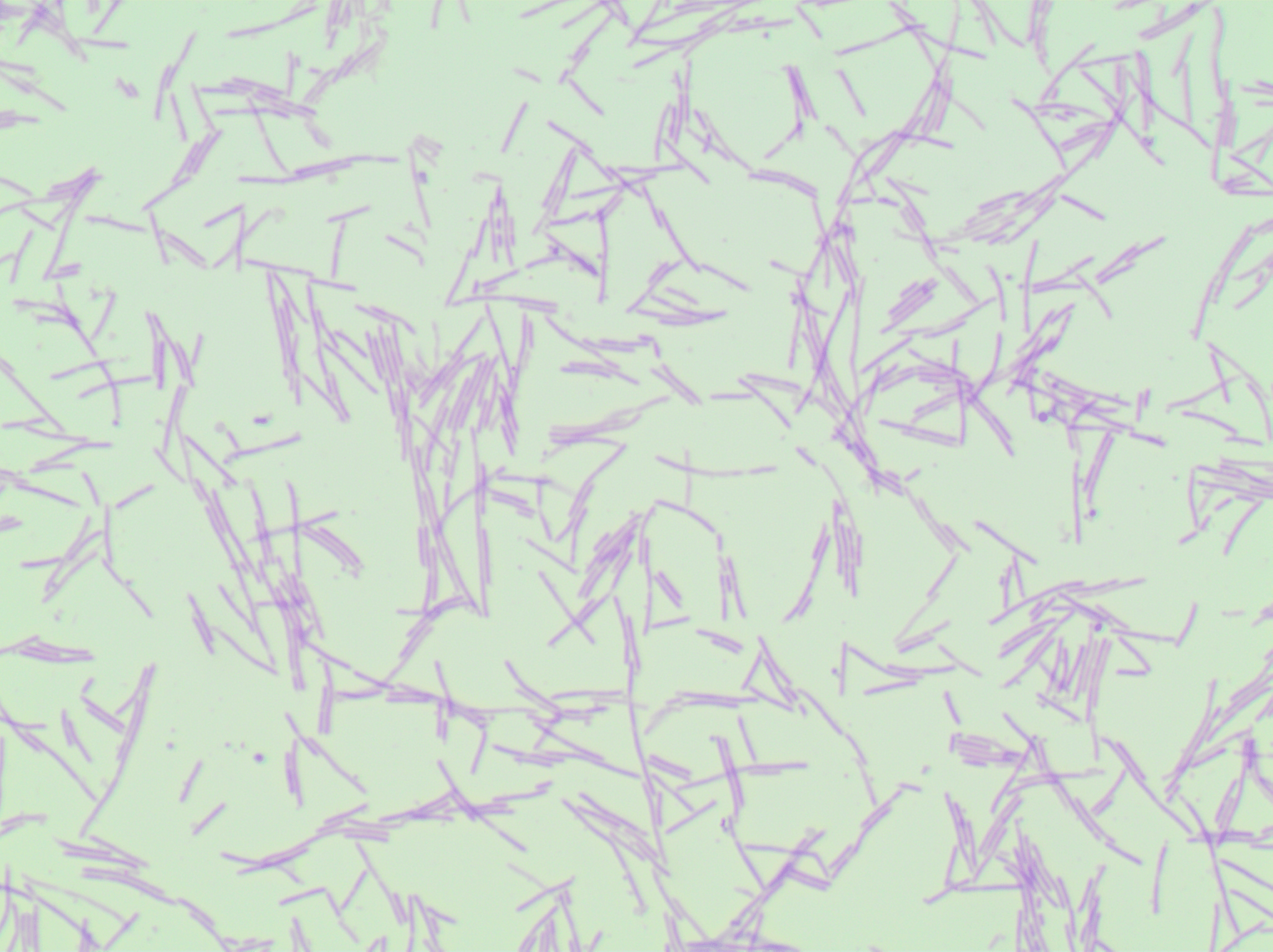}}
    \\
    
    \textbf{Lactobacillus} & \textbf{Lactobacillus} & \textbf{Lactobacillus} & \textbf{Lactobacillus}\\
    \textbf{ johnsonii} &\textbf{ johnsonii gt} &\textbf{ rhamnosus label}
    &\textbf{ rhamnosus}\\

    \frame{\includegraphics[width=1.2in]{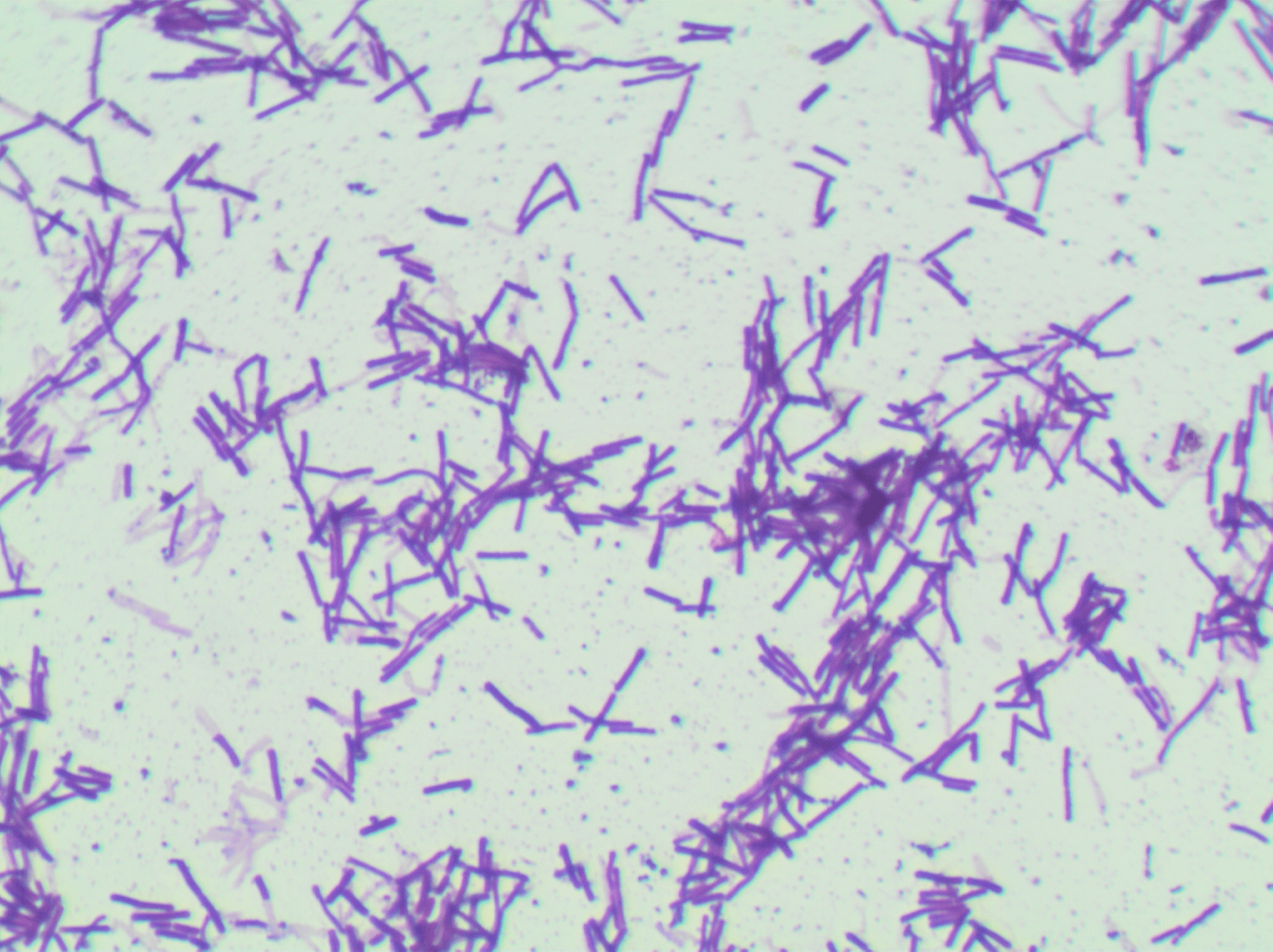}} \hfill \hfill&

    \frame{\includegraphics[width=1.2in]{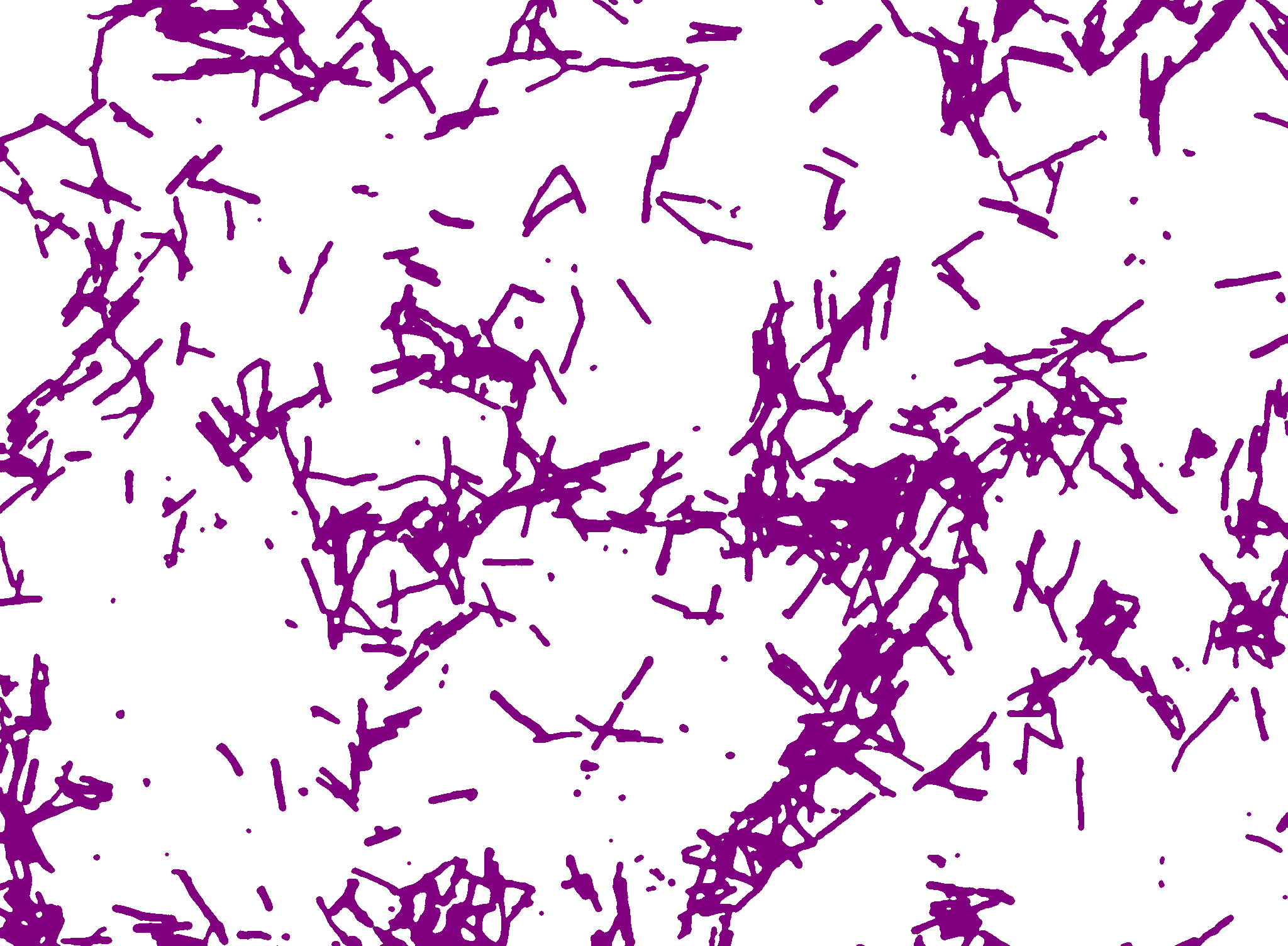}} \hfill \hfill
    &\frame{\includegraphics[width=1.2in]{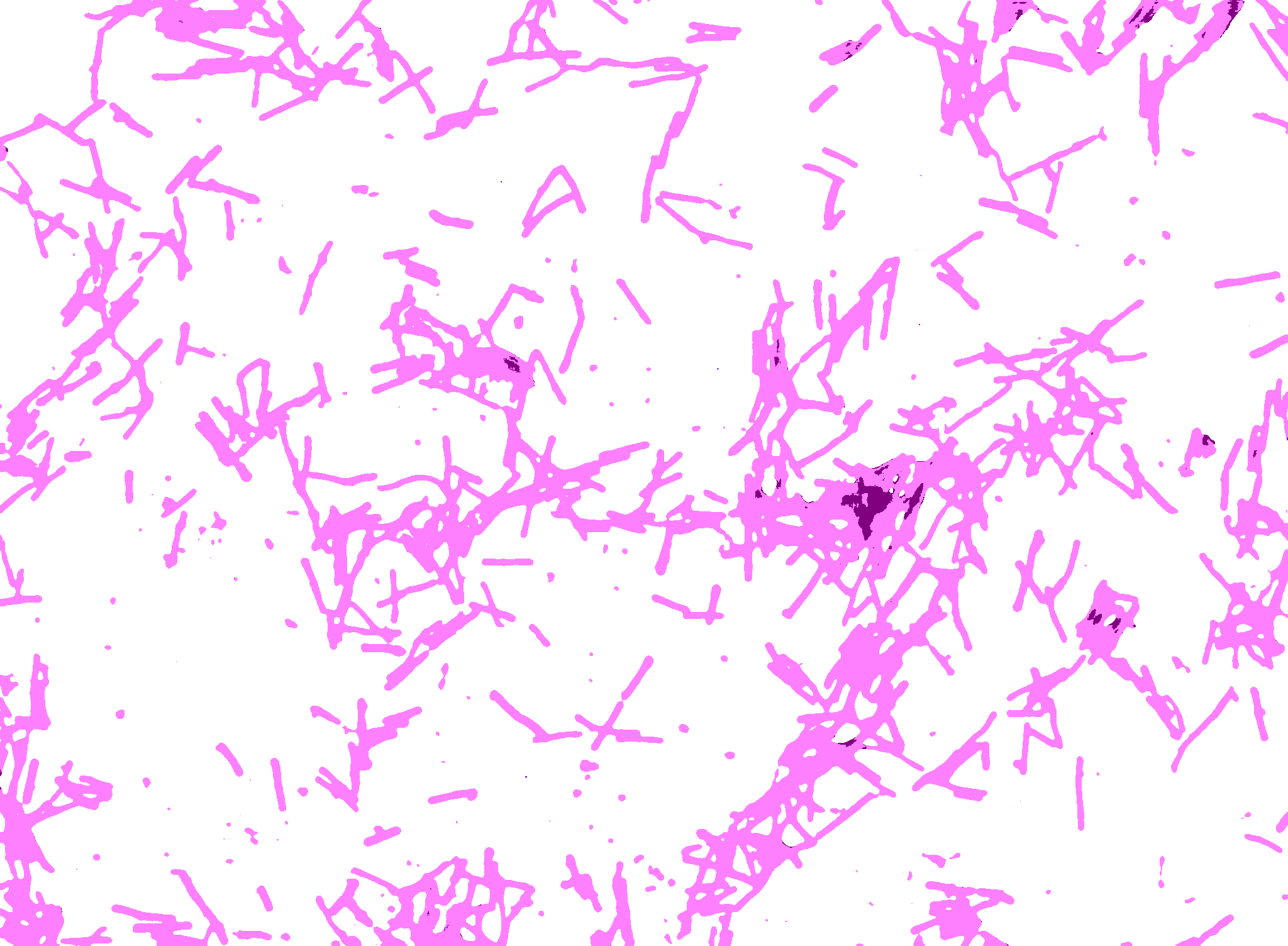}}
    &\frame{\includegraphics[width=1.2in]{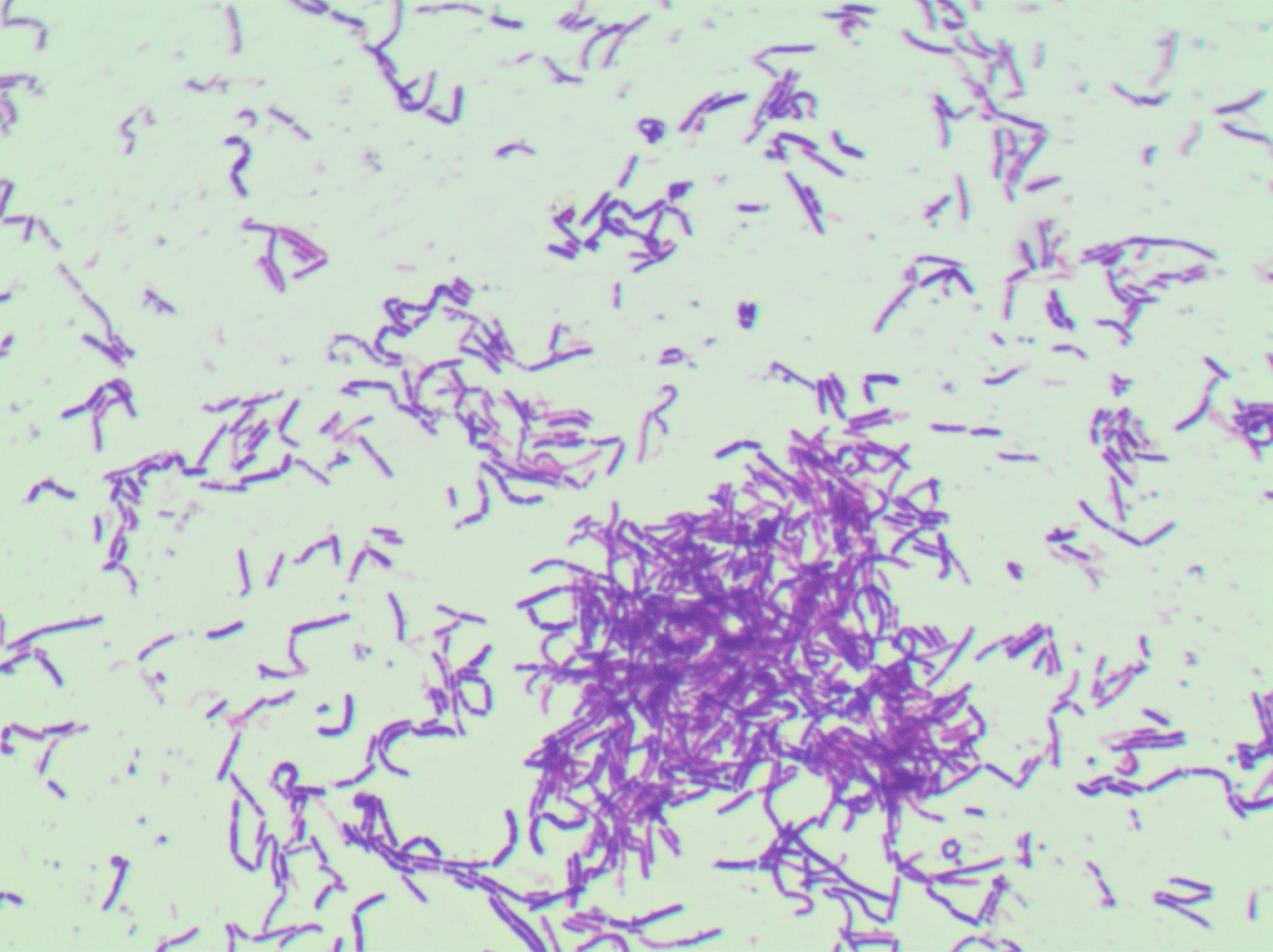}}
    \\
    
    \textbf{Lactobacillus} & \textbf{Lactobacillus} & \textbf{Lactobacillus } &\textbf{Lactobacillus }\\
    
    \textbf{reuteri} &\textbf{reuteri gt} &\textbf{jehnsenii label} &\textbf{jehnsenii}\\

    \frame{\includegraphics[width=1.2in]{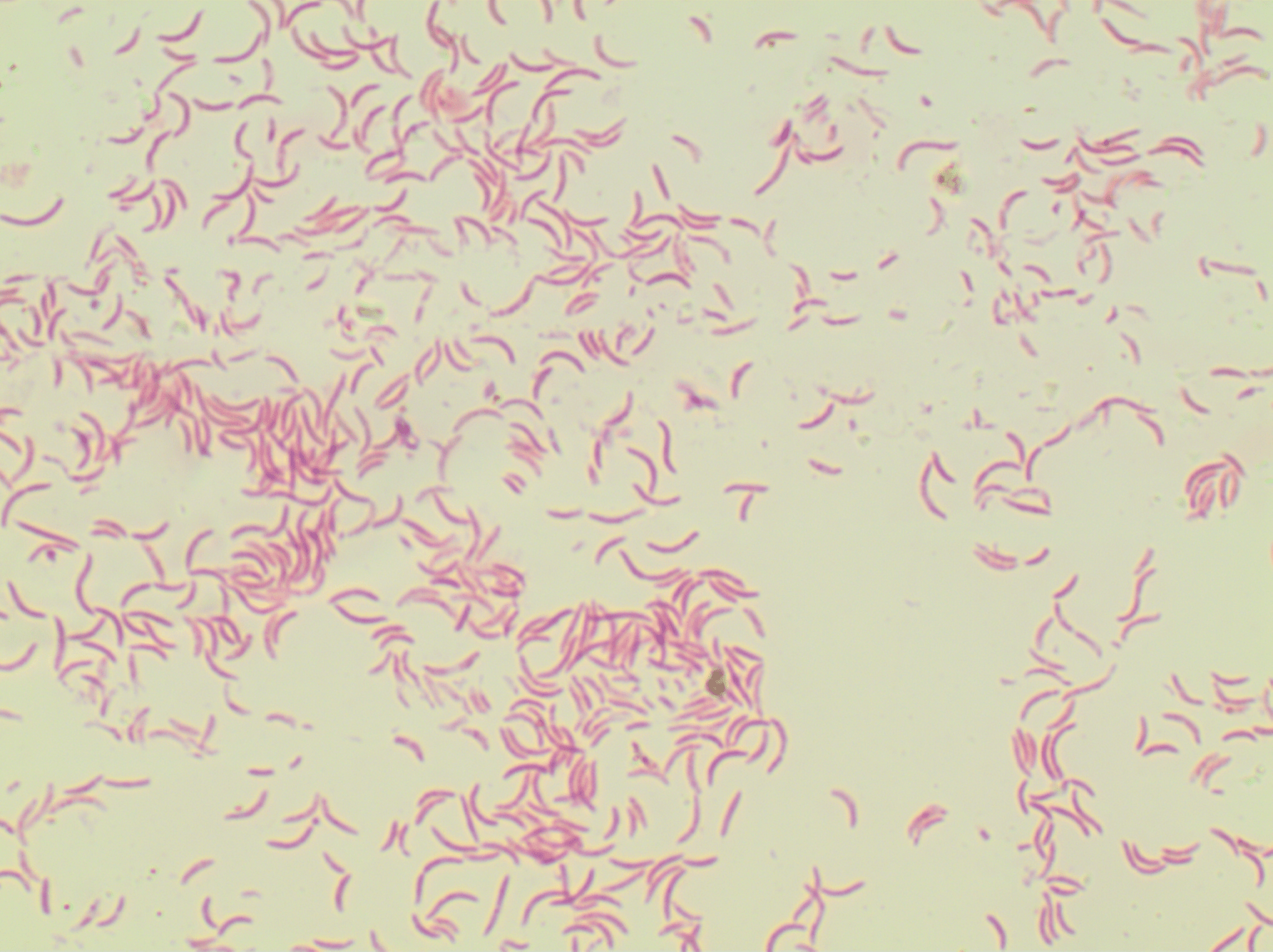}} \hfill \hfill&

    \frame{\includegraphics[width=1.2in]{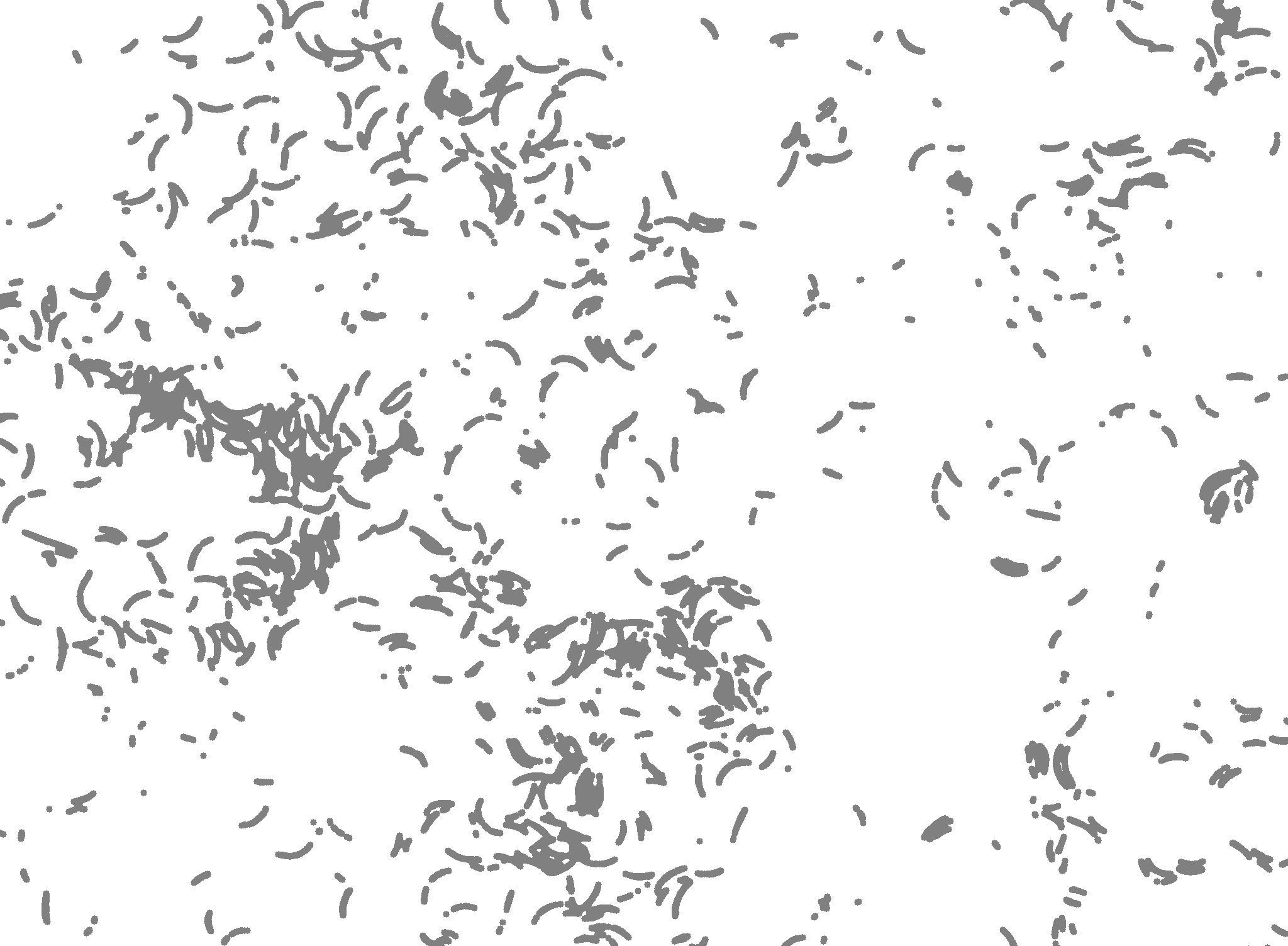}} \hfill \hfill
    &\frame{\includegraphics[width=1.2in]{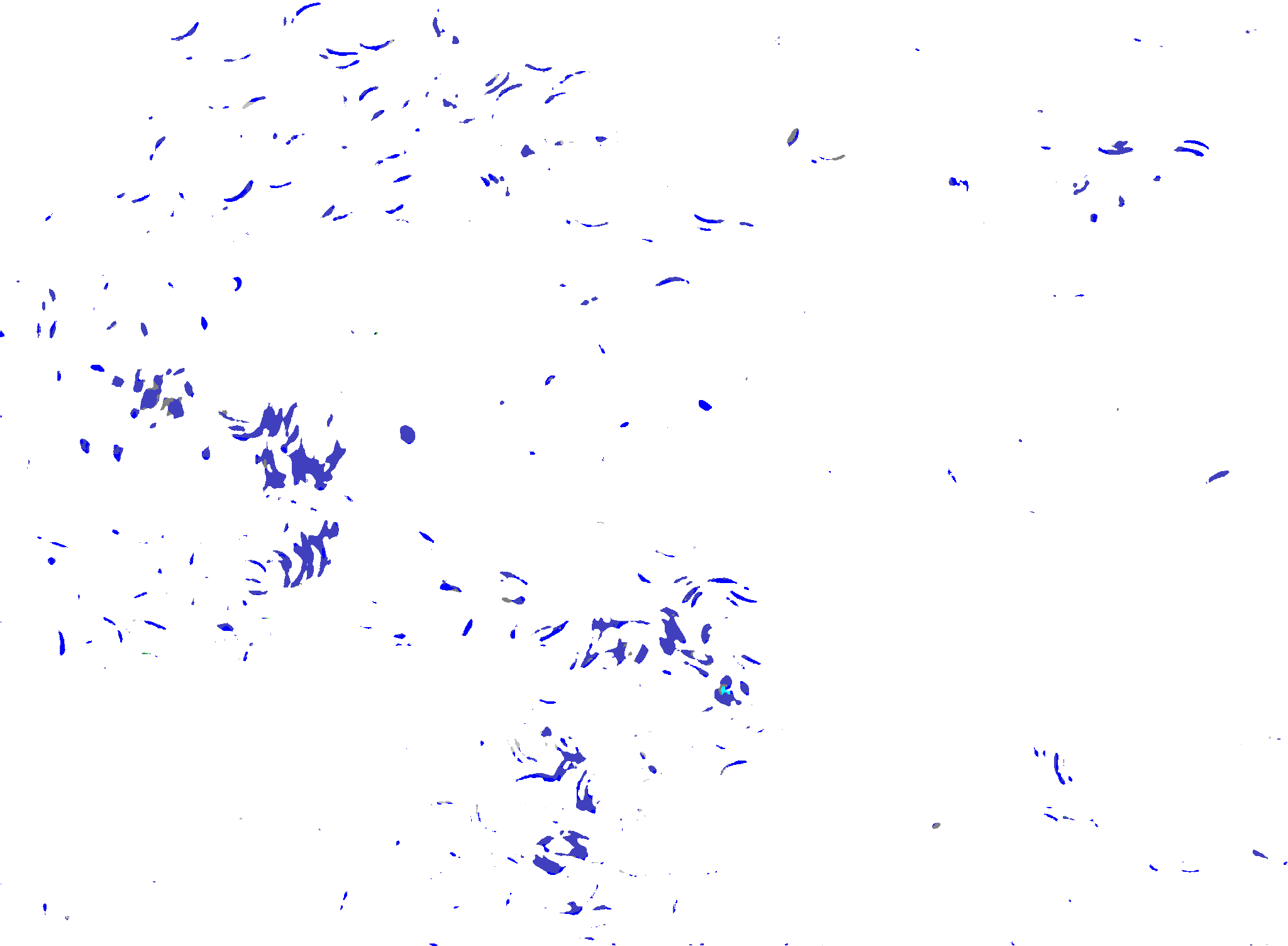}}
    &\frame{\includegraphics[width=1.2in]{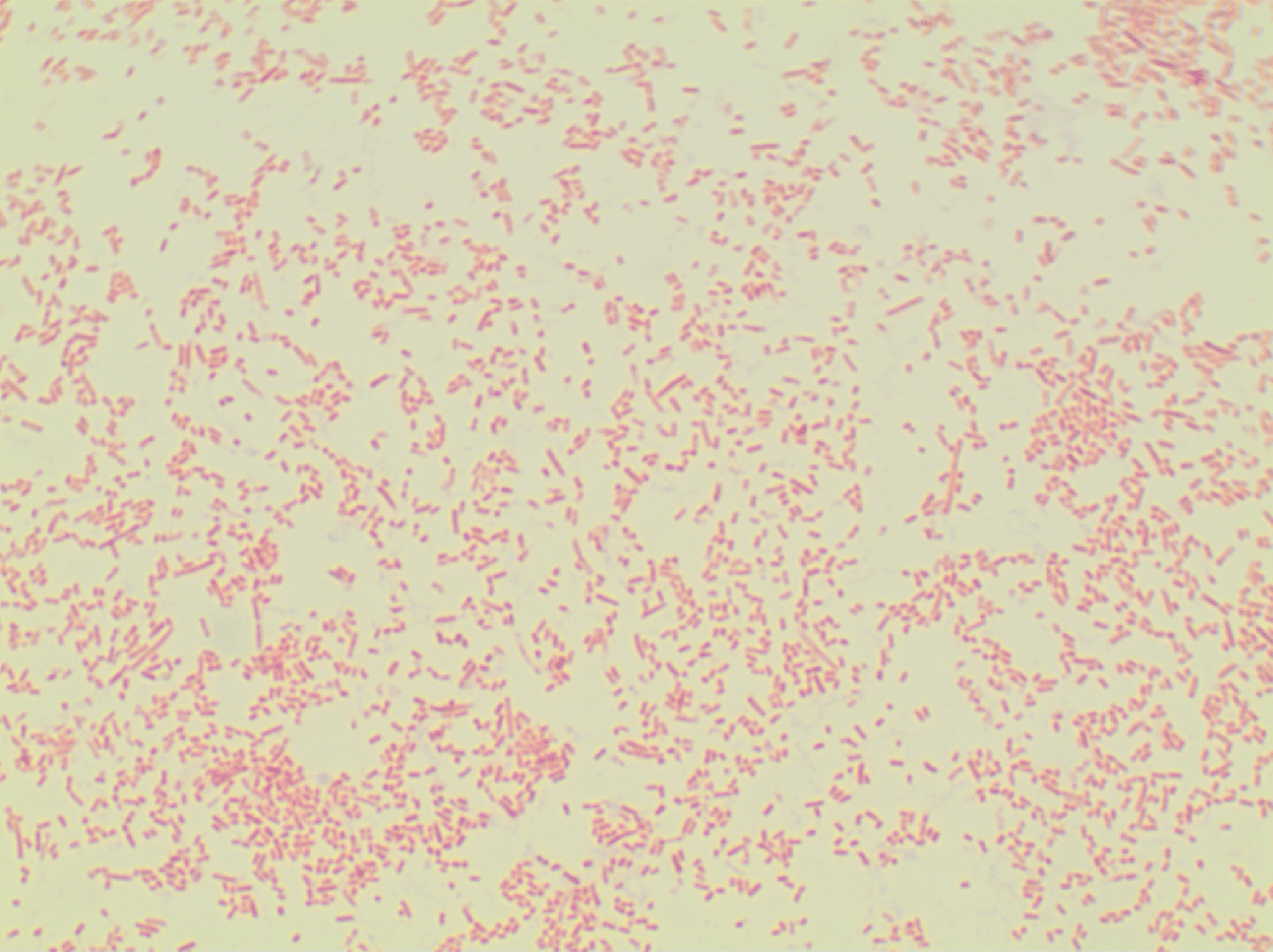}}
    \\
    
    \textbf{Porfyromonas } & \textbf{Porfyromonas} & \textbf{Proteus label} &\textbf{Proteus}\\
    
    \textbf{gingivalis} &\textbf{gingivalis gt} & &\\

    \frame{\includegraphics[width=1.2in]{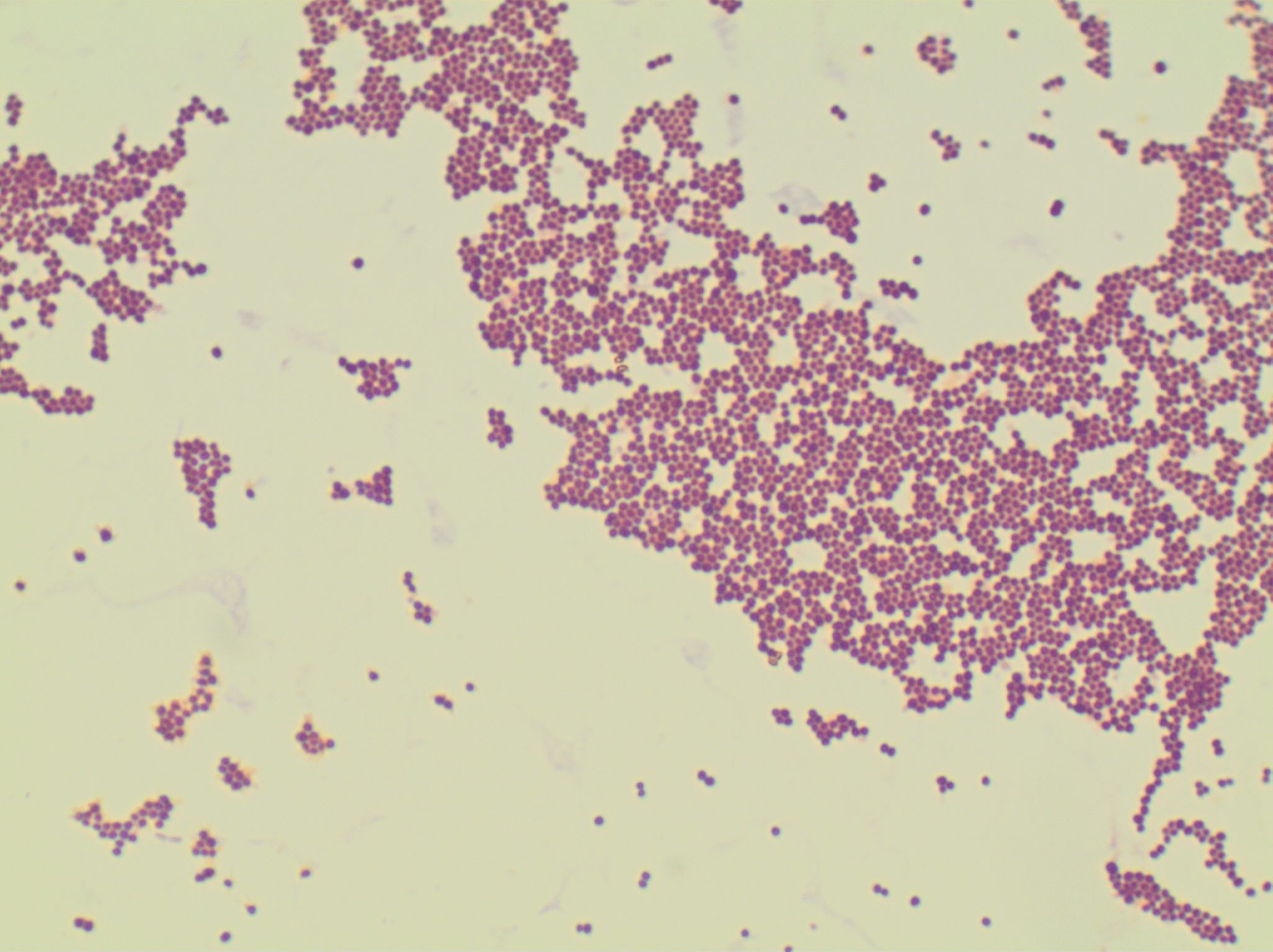}} \hfill \hfill&

    \frame{\includegraphics[width=1.2in]{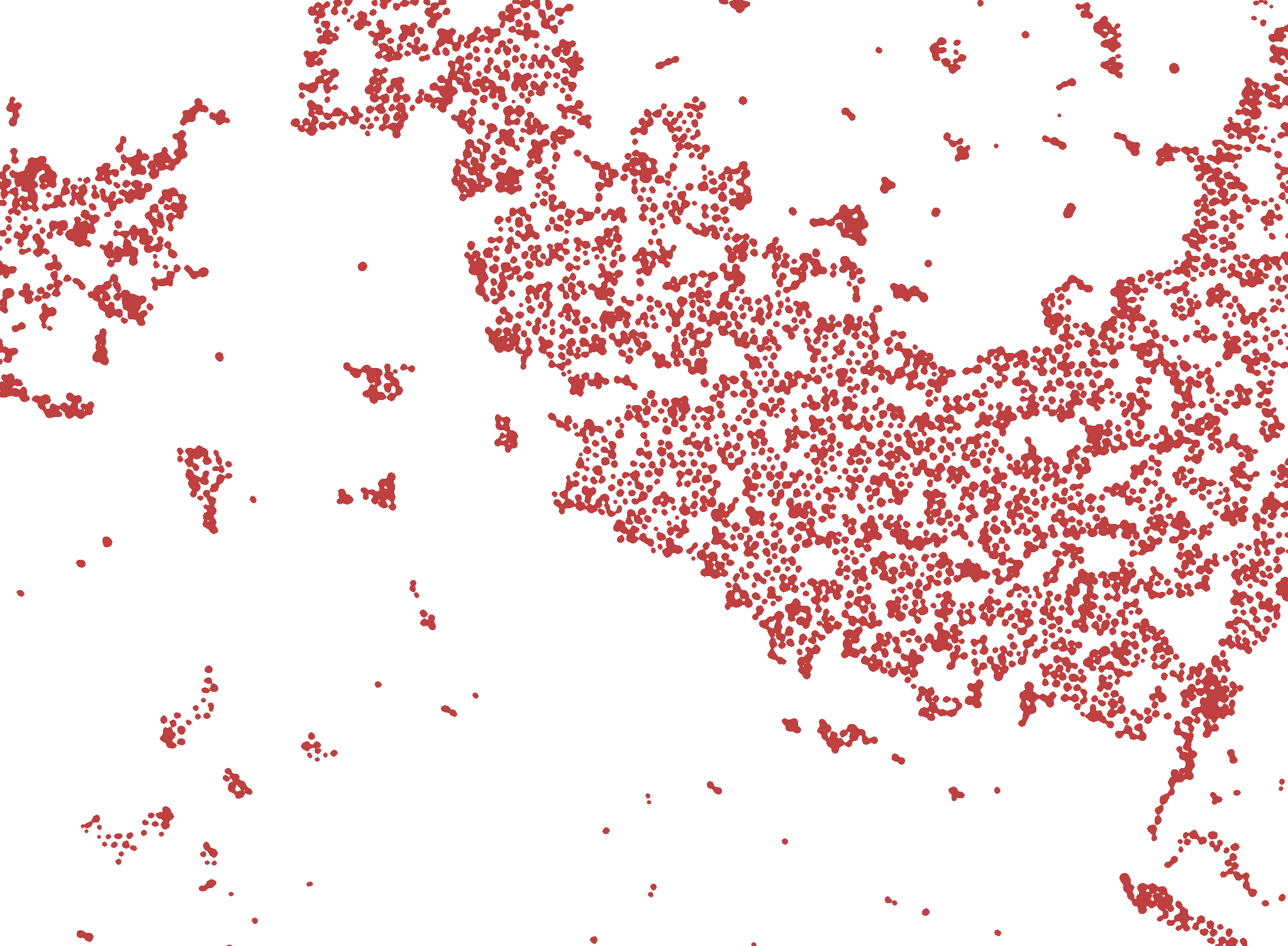}} \hfill \hfill
    &\frame{\includegraphics[width=1.2in]{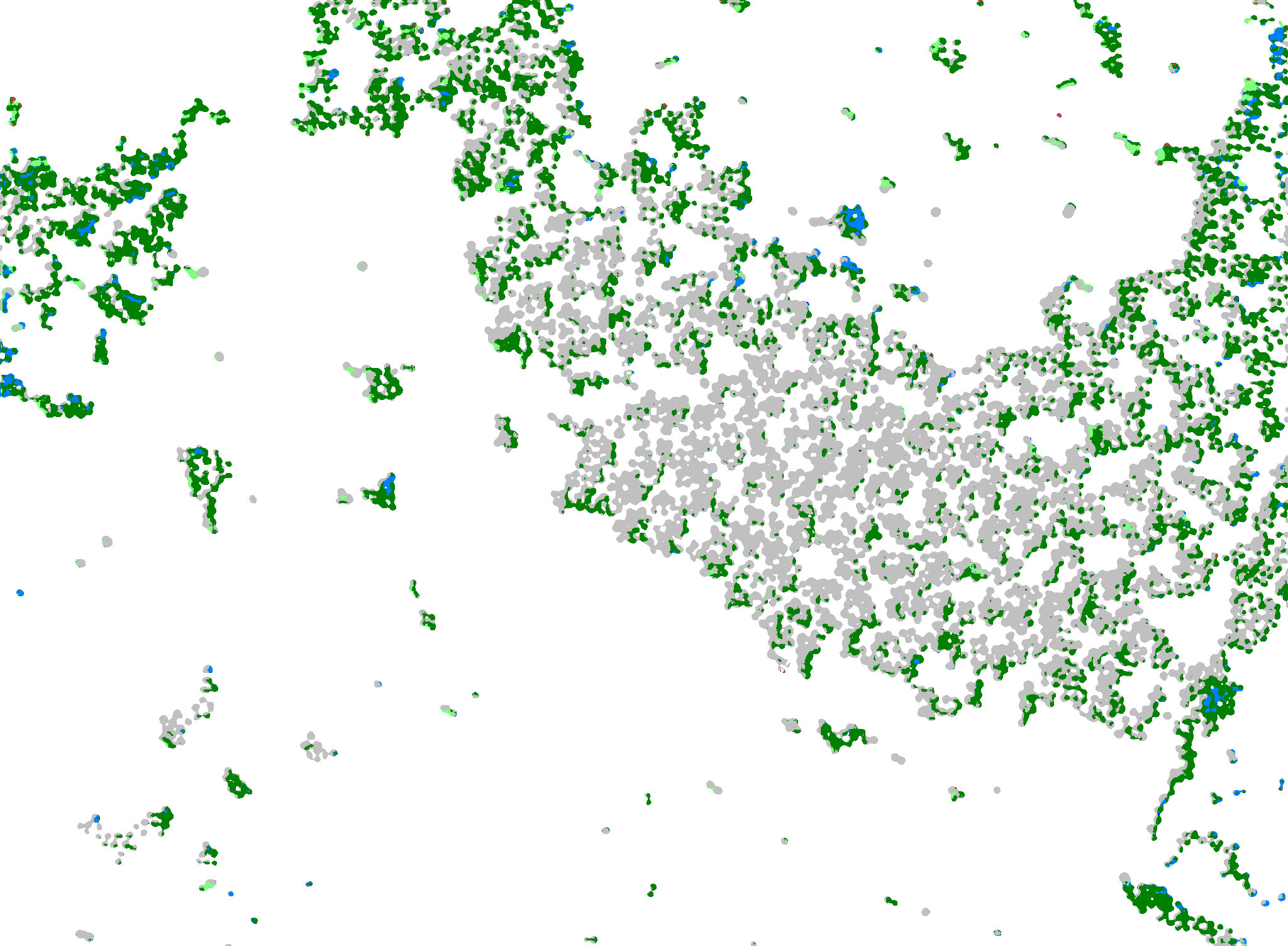}}
    &\frame{\includegraphics[width=1.2in]{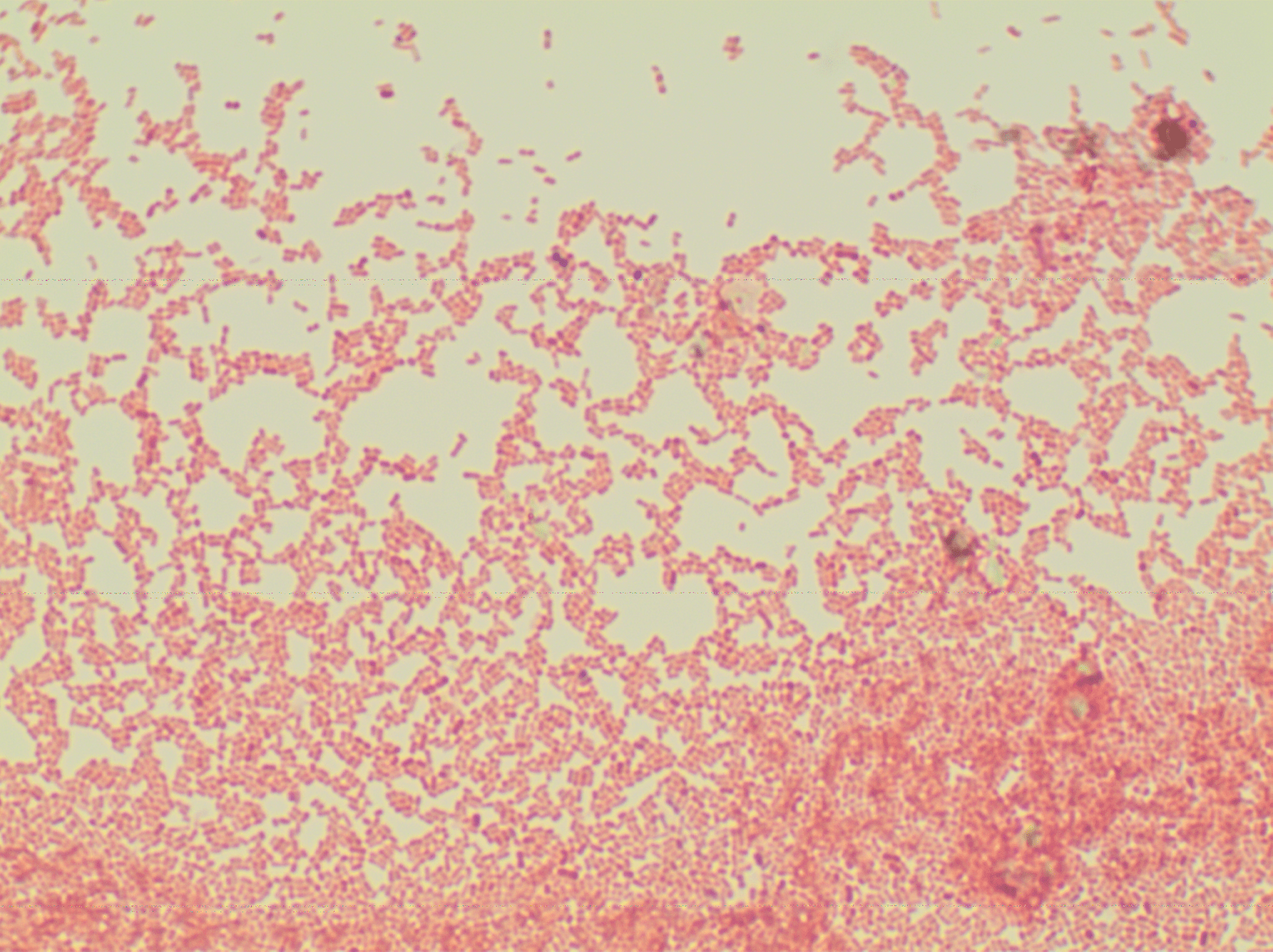}}
    \\
    
    \textbf{Staphylococcus } & \textbf{Staphylococcus} & \textbf{Acinetobacter} & \textbf{Acinetobacter}\\
    \textbf{Epidermidis} &\textbf{Epidermidis gt} &\textbf{baumanii label} &\textbf{baumanii}\\
    
  \end{tabular}
 \caption{\textbf{Examples of misclassified bacterial species}}
 \label{predictions1}
\end{figure}

 From the table it is seen that the F1 scores corresponding to Enterrococcus faecelis, Escherichia coli, Lactobacillus johnsonii and reuteri, Porfyromonas gingivalis and staphylococcus epidermidis are less than 50\%. There is similarity in the shape and arrangement of the bacteria belonging to different species. The model also gets influenced by intra-species similarity. Therefore, the model has misclassified these species of bacteria. The sample images of misclassified bacteria species are depicted in Figure \ref{predictions1}.

\section{Conclusion}
A deep-learning model for semantic segmentation and classification of bacteria genera and species from DIBaS dataset has been implemented. The images from the dataset are annotated using k-means clustering and otsu thresholding. Morphological closing is performed using circular kernels of varying sizes. Semantic segmentation is performed by ResUNet++ model. An average classification accuracy of 95\%, IoU score of 64\% and F1 score of 77\% is achieved. To the best of the authors knowledge, semantic segmentation of bacteria from microscopic images is being attempted for the first time.  We aim to use this model as a baseline to perform classification, segmentation of bacteria for other biomedical applications such as gram-stained urine smear images, sputum-smear images and so on.

\newpage

%
%
\bibliographystyle{splncs04}
\bibliography{Semiauto}
\end{document}